\documentclass[3p,preprint,12pt,margin=1in]{elsarticle}
\makeatletter\if@twocolumn\PassOptionsToPackage{switch}{lineno}\else\fi\makeatother

      \makeatletter
\usepackage{wrapfig}
\newcounter{aubio}

\usepackage[T1]{fontenc}
\usepackage[utf8]{inputenc}
\usepackage{lmodern}
\usepackage{booktabs}
\usepackage{array}
\usepackage{xcolor}
\usepackage{mdframed}
\usepackage{enumitem}
\usepackage{graphicx}
\usepackage{hyperref}
\usepackage{natbib}
\usepackage{stmaryrd}
\usepackage{booktabs}

\long\def\bioItem{%
\@ifnextchar[{\@bioItem}{\@@bioItem}}

\long\def\@bioItem[#1]#2#3{
 \stepcounter{aubio}
 \expandafter\gdef\csname authorImage\theaubio\endcsname{#1}
 \expandafter\gdef\csname authorName\theaubio\endcsname{#2}
 \expandafter\gdef\csname authorDetails\theaubio\endcsname{#3}
}

\long\def\@@bioItem#1#2{
 \stepcounter{aubio}
 \expandafter\gdef\csname authorName\theaubio\endcsname{#1}
 \expandafter\gdef\csname authorDetails\theaubio\endcsname{#2}
}

\newcommand{\checkheight}[1]{%
  \par \penalty-100\begingroup%
  \setbox8=\hbox{#1}%
  \setlength{\dimen@}{\ht8}%
  \dimen@ii\pagegoal \advance\dimen@ii-\pagetotal
  \ifdim \dimen@>\dimen@ii
    \break
  \fi\endgroup}

\def\printBio{%
  \@tempcnta=0
   \loop
     \advance \@tempcnta by 1
     \def\aubioCnt{\the\@tempcnta}
     \setlength{\intextsep}{0pt}%
     \setlength{\columnsep}{10pt}%
     \newbox\boxa%
     \setbox\boxa\vbox{\csname authorDetails\aubioCnt\endcsname}
     \expandafter\ifx\csname authorImage\aubioCnt\endcsname\relax%
      \else%
       \checkheight{\includegraphics[height=1.25in,width=1in,keepaspectratio]{\csname authorImage\aubioCnt\endcsname}}
        \begin{wrapfigure}{l}{25mm}
         \includegraphics[height=1.25in,width=1in,keepaspectratio]{\csname authorImage\aubioCnt\endcsname}
        \end{wrapfigure}\par
      \fi
     {\parindent0pt\textbf{\csname authorName\aubioCnt\endcsname}\csname authorDetails\aubioCnt\endcsname \par\bigskip%
     \expandafter\ifx\csname authorImage\aubioCnt\endcsname\relax\else%
      \ifdim\the\ht\boxa < 90pt\vskip\dimexpr(90pt -\the\ht\boxa-1pc)\fi%
     \fi}
      \ifnum\@tempcnta < \theaubio
   \repeat
   }

\makeatother

\usepackage{tabulary,xcolor}
\usepackage{amsfonts,amsmath,amssymb}
\usepackage[T1]{fontenc}
\makeatletter
\let\save@ps@pprintTitle\ps@pprintTitle
\def\ps@pprintTitle{\save@ps@pprintTitle\gdef\@oddfoot{\footnotesize\itshape \null\hfill\today}}
\def\hlinewd#1{%
  \noalign{\ifnum0=`}\fi\hrule \@height #1%
  \futurelet\reserved@a\@xhline}

\AtBeginDocument{\ifNAT@numbers \biboptions{sort&compress}\fi}

\makeatother

\usepackage{ifluatex}
\ifluatex
\usepackage{fontspec}
\defaultfontfeatures{Ligatures=TeX}
\usepackage[]{unicode-math}
\unimathsetup{math-style=TeX}
\else 
\usepackage[utf8]{inputenc}
\fi 
\ifluatex\else\usepackage{stmaryrd}\fi

\usepackage{amsthm}
\newtheorem{remark}{Remark}

\usepackage{url,multirow,morefloats,floatflt,cancel,tfrupee}
\makeatletter

\AtBeginDocument{\@ifpackageloaded{textcomp}{}{\usepackage{textcomp}}}
\makeatother
\usepackage{colortbl}
\usepackage{xcolor}
\usepackage{pifont}
\usepackage[nointegrals]{wasysym}
\usepackage{float}
\makeatletter

\usepackage{booktabs}
\usepackage{tabularx}
\usepackage{tcolorbox}

\def\mcWidth#1{\csname TY@F#1\endcsname+\tabcolsep}

\def\cAlignHack{\rightskip\@flushglue\leftskip\@flushglue\parindent\z@\parfillskip\z@skip}
\def\rAlignHack{\rightskip\z@skip\leftskip\@flushglue \parindent\z@\parfillskip\z@skip}

\@ifundefined{etal}{}{}

\usepackage{ifxetex}
\ifxetex\else\if@twocolumn\@ifpackageloaded{stfloats}{}{\usepackage{dblfloatfix}}\fi\fi

\AtBeginDocument{
\expandafter\ifx\csname eqalign\endcsname\relax
\def\eqalign#1{\null\vcenter{\def\\{\cr}\openup\jot\m@th
  \ialign{\strut$\displaystyle{##}$\hfil&$\displaystyle{{}##}$\hfil
      \crcr#1\crcr}}\,}
\fi
}

\AtBeginDocument{%
  \@ifpackageloaded{endfloat}%
   {\renewcommand\efloat@iwrite[1]{\immediate\expandafter\protected@write\csname efloat@post#1\endcsname{}}}{\newif\ifefloat@tables}%
}%

\def\BreakURLText#1{\@tfor\brk@tempa:=#1\do{\brk@tempa\hskip0pt}}
\let\lt=<
\let\gt=>
\def\processVert{\ifmmode|\else\textbar\fi}

\@ifundefined{subparagraph}{
\def\subparagraph{\@startsection{paragraph}{5}{2\parindent}{0ex plus 0.1ex minus 0.1ex}%
{0ex}{\normalfont\small\itshape}}%
}{}

\newcommand\role[1]{\unskip}
\newcommand\aucollab[1]{\unskip}
  
\@ifundefined{tsGraphicsScaleX}{\gdef\tsGraphicsScaleX{1}}{}
\@ifundefined{tsGraphicsScaleY}{\gdef\tsGraphicsScaleY{.9}}{}
\def\checkGraphicsWidth{\ifdim\Gin@nat@width>\linewidth
	\tsGraphicsScaleX\linewidth\else\Gin@nat@width\fi}

\def\checkGraphicsHeight{\ifdim\Gin@nat@height>.9\textheight
	\tsGraphicsScaleY\textheight\else\Gin@nat@height\fi}

\def\fixFloatSize#1{}
\let\ts@includegraphics\includegraphics

\def\inlinegraphic[#1]#2{{\edef\@tempa{#1}\edef\baseline@shift{\ifx\@tempa\@empty0\else#1\fi}\edef\tempZ{\the\numexpr(\numexpr(\baseline@shift*\f@size/100))}\protect\raisebox{\tempZ pt}{\ts@includegraphics{#2}}}}

\AtBeginDocument{\def\includegraphics{\@ifnextchar[{\ts@includegraphics}{\ts@includegraphics[width=\checkGraphicsWidth,height=\checkGraphicsHeight,keepaspectratio]}}}

\DeclareMathAlphabet{\mathpzc}{OT1}{pzc}{m}{it}

\def\URL#1#2{\@ifundefined{href}{#2}{\href{#1}{#2}}}

\def\UrlOrds{\do\*\do\-\do\~\do\'\do\"\do\-}%
\g@addto@macro{\UrlBreaks}{\UrlOrds}

\edef\fntEncoding{\f@encoding}

\makeatother

\newif\ifmultipleabstract\multipleabstractfalse%
    \makeatletter
\def\ead{\@ifnextchar[{\@uad}{\@ead}}
\gdef\@ead#1{\bgroup
   \def\_{\string\underscorechar\space}
   \def\{{\string\lbracechar\space}
   \def\textdagger{\string\textdagger\space}
   \def\texttildeapprox{\string\texttildeapprox\space}
   \def~{\hashchar\space}
   \def\}{\string\rbracechar\space}
   \edef\tmp{\the\@eadauthor}
   \immediate\write\@auxout{\string\emailauthor
     {#1}{\expandafter\strip@prefix\meaning\tmp}}
  \egroup
}
\gdef\emailauthor#1#2{\stepcounter{ead}
      \g@addto@macro\@elseads{\raggedright
      \let\corref\@gobble
      \eadsep\texttt{#1} (#2)
      \def\eadsep{\unskip,\space}}
}

\makeatother
  
\let\citep\cite
\let\citet\cite    
  
\usepackage{float}

\newcommand{\texttildeapprox}{{\fontfamily{pcr}\selectfont\texttildelow}}

\begin{document}

\begin{frontmatter}
    \title{
  Sustainability-Constrained Workload Orchestration for Sovereign AI Infrastructure: A Joint Compute--Network Optimization Framework}
    
\author[ciena]{Sergio Cruzes\fnref{disclaimer}}

\address[ciena]{Optical Network Engineering, Ciena Brazil,
Ciena, Av. das Na\c{c}\~{o}es Unidas, 14.171 -- 15\textordmasculine{}
andar -- Marble Tower -- Salas 1563/1564,
S\~{a}o Paulo, 04794-000, SP, Brazil}

\fntext[disclaimer]{The views, analyses, and conclusions expressed
in this paper are those of the author alone and do not represent
the position, policy, or endorsement of Ciena Corporation or any
of its affiliates. This research was conducted by the author
independently in a personal academic capacity.}
  
\begin{abstract}
AI infrastructure has transitioned from a software-centric paradigm to a system tightly bound by physical and environmental limits.  Energy availability, cooling capacity, and network connectivity now impose hard operational boundaries that cannot be relaxed through software optimization alone. This paper proposes a sustainability-constrained orchestration framework that treats carbon intensity, water usage, and power capacity as strict feasibility constraints rather than tunable penalties, and that jointly optimizes compute placement and optical network routing in a single closed-loop system. We introduce the \textbf{Feasible Sovereign Operating Region (FSOR)}---a conceptual and operational construct that characterizes the set of workloads a given infrastructure can actually sustain under its physical and regulatory endowment. Scenario-based analysis demonstrates that joint optimization yields lower environmental impact relative to baseline formulations.  Infeasibility events, rather than being optimizer failures, constitute precise, telemetry-grounded signals that sovereign AI operation requires infrastructure investment or workload reduction.

\end{abstract}

    
  \end{frontmatter}

\section{Introduction}
\label{sec:intro}
 
AI infrastructure has undergone a fundamental shift. For most of its history, progress in artificial intelligence was primarily driven by software: better algorithms, improved training procedures, and more efficient model architectures. Physical infrastructure was largely a background concern, scalable by adding commodity hardware. That paradigm has changed. The emergence of large-scale foundation models and the rapid expansion of AI deployment have transformed infrastructure from a supporting layer into a primary constraint~\cite{iea2025energyai,mytton2025ecology}.

Today, AI workloads at scale impose demands that directly engage physical limits. Power consumption at major AI data centers reaches levels that strain grid interconnections and require dedicated energy agreements~\cite{iea2025energyai}. Cooling requirements have outpaced the capacity of conventional air-cooling systems, driving a transition to liquid and immersion cooling with corresponding water demands~\cite{Mytton2021}. The growth of AI electricity consumption has become a visible factor in national energy planning. These are not merely engineering challenges to be resolved through optimization---they define hard feasibility boundaries on where AI systems can be built, how large they can grow, and under what regulatory conditions they can operate.

A data center that exceeds its local carbon budget cannot expand. A facility that exhausts its water permit cannot operate at full capacity during peak summer months. A site that saturates its grid interconnection cannot accept new workloads, regardless of compute availability or capital investment. In this context, sustainability functions as a \emph{feasibility constraint} rather than an efficiency target. The distinction is significant: an efficiency target can always be traded against other objectives; a feasibility constraint cannot. This paper formalizes that distinction and builds it into the design of an orchestration framework.

AI infrastructure decisions are often treated as purely technical optimization problems, but the technology is not neutral: its impacts---and the capacity to deploy it at all---depend on regulatory frameworks, grid conditions, and institutional context~\cite{ye2026energyjustice}. Software-level openness, while valuable for reducing vendor lock-in, does not eliminate dependence on underlying compute, network, and energy infrastructure; operational autonomy requires physical feasibility, not only licensing freedom. In energy systems, AI-driven efficiency gains are real but unevenly distributed across regions and social groups, reinforcing the need to treat infrastructure constraints not only as engineering limits but as structural determinants of who can participate in AI-driven systems. Sustainability-aware orchestration must therefore extend beyond compute scheduling to encompass coordinated management of multi-energy infrastructure---on-site generation, storage, and thermal systems---making AI infrastructure optimization an inherently cross-domain problem~\cite{nkwawir2025multienergy}.

\subsection{Limitations of existing approaches}

The dominant thread in sustainable computing addresses carbon-aware scheduling: shifting flexible workloads in time toward low-carbon grid periods, or in space toward low-carbon regions~\cite{radovanovic2021carbon,cote2025locational}.
This approach has demonstrated meaningful reductions in grid emissions, but it operates within three systematic limitations that this work addresses.

First, sustainability is typically modeled as an optimization objective---a term to be minimized subject to performance constraints. When it conflicts sharply with performance or cost, the optimizer trades it away. This is appropriate for \emph{preference} targets but inappropriate when the sustainability limit is a regulatory or physical absolute.  A penalty term cannot enforce a legal maximum; only a hard constraint can.

Second, existing frameworks decouple compute placement from network routing. Placement models assume that traffic can be routed between any two sites at negligible cost.  Routing models treat compute demand as a fixed external input. In practice, the two decisions interact: placing a workload at a low-carbon site may require routing its traffic over a high-latency optical path that violates service requirements~\cite{singla2014}. Solving either problem in isolation cannot find the globally feasible solution~\cite{cottier2025advances}.

Third, sustainability parameters are typically evaluated retrospectively, using historical or batch data. Grid carbon intensity varies on sub-hourly timescales. Water availability varies seasonally. Grid capacity fluctuates with load and generation mix. A framework that makes placement decisions on stale data may violate constraints that were not visible at planning time~\cite{cruzes2026telemetry}.

While prior work demonstrates that digitalization can improve sustainability outcomes, this paper shows that such improvements are fundamentally bounded by infrastructure feasibility constraints.

\subsection{This work}

This paper moves beyond optimization-based sustainability by showing that AI infrastructure is fundamentally constrained by physical feasibility. We introduce the Feasible Sovereign Operating Region (FSOR), which defines the set of workloads that can be operated under joint compute, network, and sustainability constraints. This reframes sustainability from an objective to be optimized into a boundary that determines whether operation is possible.

This paper addresses all three limitations through a sustainability-constrained orchestration framework with three defining properties. First, sustainability limits are modeled as \emph{hard constraints}: carbon and water thresholds eliminate infeasible configurations from the solution space entirely, rather than penalizing them in the objective. Second, compute placement and optical network routing are solved \emph{jointly} in a single optimization, making the coupling between placement decisions and network costs explicit.  Third, the optimization operates in a \emph{closed loop} driven by real-time telemetry and validated against a digital twin before any configuration change is applied.

A central analytical contribution is the \textbf{Feasible Sovereign Operating Region (FSOR)}: the set of workload configurations for which the joint optimization has at least one valid solution under a given infrastructure endowment and policy regime. The FSOR makes sovereign AI capacity measurable in physical terms. When a region's FSOR shrinks to the empty set, that region cannot sustain AI operation at the required constraint levels---independently of software capability or compute investment.

The framework is grounded in the infrastructure-sovereignty perspective developed in~\cite{cruzes2026sovereignty}, which positions sustainability as a first-order design constraint that shapes a region's capacity to deploy and evolve AI systems over time. The present paper operationalizes that conceptual framework into a concrete optimization formulation and an implementable control architecture.

\subsection{Contributions}

The specific contributions of this paper are:
\begin{itemize}
  \item A formal optimization model for sustainability-constrained workload
      placement in which carbon intensity and water usage are hard feasibility
      boundaries, together with the \textbf{FSOR} construct that characterizes
      the resulting feasible space as a function of measurable physical and
      regulatory parameters (Section~\ref{sec:problem}).
  \item A joint compute--network formulation that couples workload assignment
        with optical flow routing under latency and capacity constraints
        (Section~\ref{sec:problem}).
    \item An agentic closed-loop control architecture integrating streaming
        telemetry, optimization, and digital twin validation
        (Section~\ref{sec:architecture}).
  \item Scenario-based analysis demonstrating the qualitative and comparative
        advantage of joint optimization under realistic infrastructure
        conditions (Section~\ref{sec:evaluation}).
  \item A sovereignty analysis mapping infrastructure endowments to FSOR boundaries (Section~\ref{sec:sovereignty}).
\end{itemize}

\section{Related Work}
\label{sec:related}
 
\subsection{Carbon-aware scheduling}

The foundational work on carbon-intelligent computing was introduced by Radovanovic et al.~\cite{radovanovic2021carbon}, who demonstrated that shifting delay-tolerant batch jobs toward low-carbon grid periods reduces net emissions at Google-scale deployments without degrading throughput. Cote and Sun~\cite{cote2025locational} formalized locational marginal emissions as
grid-granularity carbon signals, enabling spatial as well as temporal workload shifting.

These contributions establish the operational feasibility of carbon-aware scheduling and provide the signal infrastructure that our framework builds on.

Recent work has further demonstrated that carbon-aware scheduling alone is insufficient to achieve truly low-carbon operation in datacenters. The Carbon Explorer framework shows that Net Zero accounting masks significant hourly carbon emissions due to the intermittent and geographically variable nature of renewable energy supply~\cite{acun2023carbonexplorer}. Achieving 24/7 carbon-free operation requires coordinated management of multiple mechanisms, including workload shifting, energy storage, and renewable energy deployment, each introducing trade-offs between operational and embodied carbon. These results highlight that carbon-aware scheduling is fundamentally constrained by energy system dynamics, reinforcing the need to treat sustainability as a feasibility condition rather than solely as an optimization objective.

A complementary line of work has quantified the inherent limits of carbon-aware spatiotemporal workload shifting in large-scale cloud systems. Sukprasert et al.~\cite{Sukprasert2024}
show that even under idealized conditions with perfect
knowledge of future carbon intensity and unrestricted
workload mobility, the achievable carbon reductions are
bounded and often significantly lower in practice due to
constraints such as limited workload flexibility, latency
requirements, and datacenter capacity. In particular,
long-running workloads, which account for a large fraction
of energy consumption, exhibit limited temporal flexibility
and therefore cannot effectively exploit low-carbon periods.
Moreover, spatial shifting benefits are constrained by
performance and regulatory requirements, and simple
migration policies already capture most of the achievable
gains. These results indicate that carbon-aware scheduling
is fundamentally limited not only by energy system dynamics
but also by workload characteristics and system-level
constraints.

However, prior work predominantly models carbon as an optimization objective—a term to be minimized subject to performance constraints—rather than as a hard feasibility limit. As a result, these approaches implicitly assume that environmentally optimal placements remain operationally achievable. In practice, this assumption breaks down under real infrastructure constraints. Limited workload flexibility, latency requirements, and capacity constraints restrict the extent to which workloads can be shifted, while low-carbon sites may only be reachable via high-latency optical paths. Consequently, compute-only formulations systematically underestimate the true feasibility cost of environmental compliance. This gap highlights the need to treat sustainability not as an optimization objective, but as a boundary condition that determines whether a given workload placement is viable.

\subsection{Green data center optimization}

Buyya et al.~\cite{buyya2013energyefficient} laid the foundation for energy-efficient resource management in cloud infrastructure, demonstrating virtual machine consolidation under power caps. Silva et
al.~\cite{silva2024decarbonization} surveyed decarbonization strategies for
high-performance computing, identifying cooling efficiency, renewable
procurement, and carbon-aware scheduling as the principal levers. Hoxha et
al.~\cite{hoxha2025llm} specifically addressed the deployment-aware problem of
carbon- and water-efficient LLM serving, proposing infrastructure-aware model
routing.

This body of work addresses the single-site or compute-layer problem. None of these contributions jointly models the optical network layer, and none treats sustainability as a hard feasibility boundary that may render certain
configurations infeasible rather than merely suboptimal.

Recent work has extended data center optimization beyond compute-level resource management to include integrated energy system coordination. Nkwawir et al.~\cite{nkwawir2025multienergy} propose a carbon-aware workload scheduling framework that jointly optimizes IT workloads and multi-energy infrastructure, including photovoltaic generation, battery storage, combined cooling, heating and power (CCHP) systems, and thermal energy storage. Their
formulation, based on mixed-integer linear programming, demonstrates significant reductions in operational emissions and energy cost through coordinated energy dispatch and workload management.

A broader review of carbon-aware scheduling techniques further reveals systematic limitations in current approaches. Recent studies show that most existing methods consider either temporal or spatial workload shifting in isolation, with only a small number of works performing true joint spatio-temporal optimization~\cite{asadov2025spatiotemporal}.
Additionally, the majority of approaches focus exclusively on
operational electricity, while embodied emissions associated with hardware production remain largely unaccounted for. The same studies highlight that effective temporal shifting requires accurate forecasting of both workload demand and carbon intensity, and that workload shiftability itself is constrained by application characteristics such as duration, deadlines, and interruptibility. Moreover, network-related costs of workload migration are rarely modeled, despite their potential impact on feasibility. These findings indicate that existing data center optimization approaches capture only partial aspects of the problem and fail to represent the full set of constraints governing sustainable operation.

Accurate forecasting of resource utilization has emerged as a critical enabler of carbon-aware data center operation. Recent work proposes hybrid deep learning models that combine transformer-based architectures with recurrent networks to capture both long-range dependencies and short-term
dynamics in multivariate workload patterns~\cite{sallam2025temposight}. These
approaches demonstrate that improvements in forecasting accuracy translate directly into more effective carbon-aware scheduling and resource provisioning.

However, these approaches remain confined to either prediction or energy-layer optimization within individual data centers. They do not incorporate cross-site optimization, optical network constraints, or latency-sensitive workload placement across geographically distributed sites. As a result, they cannot capture the interaction between forecasting, placement, and network feasibility that defines the FSOR in this work.

Recent empirical studies have identified multiple pathways through which AI influences energy systems, including improvements in energy efficiency, acceleration of green innovation, and optimization of system-level resource
allocation~\cite{ye2026energyjustice}. These mechanisms reinforce the role of AI as a driver of infrastructure efficiency, but also highlight that its system-level impact depends on how constraints and policies are enforced.

\subsection{Joint compute--network optimization}

Joint optimization of compute placement and network routing has been studied in mobile edge computing and content delivery contexts.  Xin et al.~\cite{xin2026dnn} addressed load-balanced DNN inference offloading in metro optical networks, treating latency and capacity as joint constraints.  Gu et al.~\cite{gu2020ml} provided a comprehensive survey of machine learning techniques for intelligent optical networks, covering traffic prediction and routing optimization.

Recent advances in carbon-aware scheduling have introduced
mechanisms for adapting workload placement and execution to
spatiotemporal variations in grid carbon intensity. These
approaches leverage real-time or forecasted carbon signals
to shift workloads across time and regions, improving the
carbon efficiency of data center operations~\cite{EmergentMind2025}. However, they
primarily operate at the compute or data center level and
treat sustainability as an optimization objective rather
than a hard feasibility constraint. Critically, they do not
incorporate optical network constraints, routing feasibility,
or latency-induced coupling between geographically distributed
sites. As a result, carbon-aware scheduling remains decoupled
from the network layer and cannot capture the cross-domain
interactions that arise in distributed AI infrastructure.

These works demonstrate the value of joint formulations but do not incorporate sustainability constraints.  The specific problem of sustainability-constrained joint compute--optical optimization for AI workloads---where environmental limits directly interact with routing and placement decisions---has not been previously addressed.

Recent work on carbon-aware spatiotemporal workload shifting further highlights the limitations of existing placement formulations under realistic constraints. Large-scale empirical analysis shows that while spatial workload migration can theoretically achieve substantial carbon reductions, its practical benefits are significantly constrained by latency requirements, datacenter capacity, and regulatory limitations~\cite{Sukprasert2024}. Moreover, these studies demonstrate that simple placement policies often achieve most of the attainable gains, with more sophisticated optimization strategies providing marginal improvements. Importantly, such approaches model workload placement at the cloud level but do not incorporate optical network constraints, routing feasibility, or cross-layer interactions. As a result, they overestimate the flexibility of compute placement and fail to capture the coupled feasibility problem that arises when network, latency, and sustainability constraints must be satisfied simultaneously.

\subsection{Sovereignty and infrastructure}

Fratini and Hine~\cite{fratini2024digital} analyzed digital sovereignty as a spectrum of control bounded by technical and economic realities rather than a binary property. This perspective has been reinforced by subsequent work highlighting that sovereignty depends not only on jurisdictional or regulatory authority, but on the ability to operate infrastructure under real-world constraints. Empirical studies of compute infrastructure further show that apparent sovereignty based on data location or policy can overestimate actual control, as effective autonomy depends on ownership, supply chains, and physical deployment of compute resources.

Chetty et al.~\cite{chetty2025sovereign6g} examined the implications of sovereign AI in the context of 6G network architectures, while Cruzes~\cite{cruzes2026sovereignty} introduced an infrastructure-centric sovereignty framework arguing that a region's capacity to develop and operate AI systems is determined by its physical and environmental infrastructure endowment rather than software capability alone. This framing shifts sovereignty from a legal or conceptual notion to an operational property grounded in infrastructure availability and constraints.

Recent empirical studies further reinforce this infrastructure-centric view by demonstrating that the effectiveness of carbon-aware workload orchestration is fundamentally limited by physical and system-level constraints. Even under idealized conditions, the achievable carbon reductions from spatiotemporal workload shifting are significantly reduced by latency requirements, capacity limitations, and regulatory constraints~\cite{Sukprasert2024}. These results highlight that infrastructure is not only an enabler of AI operation, but also a binding constraint that shapes the feasible set of operational strategies. As a consequence, sovereignty must be understood not only as a function of control over infrastructure, but also as a function of the constraints that govern how that infrastructure can be used in practice.

Recent empirical work on the global distribution of AI compute infrastructure further sharpens this perspective by revealing a pronounced geographic asymmetry in access to compute resources. A global census of public cloud GPU infrastructure shows that a small number of countries host the majority of
advanced AI compute capacity, while many regions possess only limited inference-oriented infrastructure or none at all~\cite{lehdonvirta2024compute}.
This has led to the characterization of a ``Compute North'', where countries host sufficient infrastructure for AI development, a ``Compute South'', where
infrastructure supports primarily deployment, and a ``Compute Desert'', where no meaningful AI compute capacity exists. These findings demonstrate that the ability to govern, develop, and operate AI systems is not uniformly distributed,
but is instead structurally constrained by the geographic availability of compute infrastructure. As a result, sovereignty is not only determined by control over infrastructure, but also by whether such infrastructure exists locally in the first place.

Recent large-scale empirical analyses of data center energy demand further quantify the magnitude of these constraints. Data center electricity consumption in the United States reached approximately 176~TWh in 2023, representing 4.4\% of total national electricity use, and is projected to increase to between 325 and 580~TWh by 2028 depending on infrastructure and deployment scenarios~\cite{shehabi2024datacenter}. This growth is driven primarily by AI workloads and accelerator-based systems, whose power and cooling requirements significantly exceed those of traditional computing infrastructure. These findings highlight that AI infrastructure expansion is directly coupled to energy system capacity, cooling technologies, and grid conditions, reinforcing that sovereignty is bounded not only by ownership or control of infrastructure, but by the physical limits of the energy systems that sustain it.

The present paper operationalizes this argument by casting AI infrastructure sovereignty as a constrained optimization and control problem, where joint compute placement and network routing must satisfy explicit sustainability limits, and by defining the Feasible Sovereign Operating Region (FSOR) as the resulting feasibility boundary.
\section{System Model and Problem Formulation}
\label{sec:problem}

\subsection{Infrastructure Model}
\label{subsec:infra_model}

We model AI infrastructure as a coupled compute--network--energy system in which workload placement, network routing, and sustainability constraints must be satisfied simultaneously. The model has three layers---compute sites, optical transport, and workloads---each with its own parameter space and temporal dynamics. We describe each in turn before defining the joint
feasibility conditions that couple them.

\subsubsection{Compute Sites}

Let $\mathcal{S} = \{s_1, \dots, s_N\}$ denote the set of $N$
geographically dispersed AI-oriented data center sites. Each site
$s_i$ is characterized at telemetry cycle $t$ by a tuple of
time-varying parameters; the three layers---compute sites, optical
transport, and workloads---are described in turn before the joint
feasibility conditions that couple them are defined.
\begin{itemize}
  \item $P_i(t)$: the maximum available electrical power
    \textnormal{[kW]}, determined by contracted grid capacity,
    on-site generation, and any active demand-response obligations;
  \item $\gamma_i(t)$: the marginal carbon intensity of the local
    grid \textnormal{[gCO$_2$eq/kWh]}, drawn from grid telemetry
    or forecast distributions;
  \item $\omega_i(t)$: the water usage effectiveness per unit of
    IT load \textnormal{[L/kWh]}, a function of cooling technology,
    ambient temperature, and humidity;
  \item $\bar{\Gamma}_i$: the policy-defined carbon intensity
    ceiling \textnormal{[gCO$_2$eq/kWh]} above which workload
    placement at $s_i$ is not permitted;
  \item $\bar{\Omega}_i(t)$: the water draw permit limit
    \textnormal{[L/h]}, which may tighten seasonally as described
    in~\S\ref{subsec:endowments}.
\end{itemize}

Power and carbon intensity are treated as rapidly varying parameters
and are updated at each telemetry cycle, with a granularity of
minutes to hours depending on grid operator data availability.
Water intensity varies more slowly---governed by ambient conditions
and cooling system dynamics---and is updated on a daily or seasonal
basis. The asymmetry in temporal resolution across parameters is not
merely a modeling convenience: it reflects the physical timescales
at which each constraint can meaningfully bind. Carbon-aware
scheduling can react to sub-hourly grid signals; water permit
compliance must be planned over days to weeks, and an operator
managing both simultaneously must therefore maintain two distinct
control loops operating at fundamentally different cadences.

Sites may also differ in their structural dependency profile. A site
operated on third-party cloud infrastructure inherits the provider's
scheduling policies and jurisdictional obligations, which may further
reduce the effective power headroom or introduce site eligibility
restrictions that are invisible in the physical parameters
above~\cite{Leenes2011,Swire2012}. We treat such external
obligations as additional upper bounds on $P_i(t)$ or as site
eligibility flags, and return to their sovereignty implications
in~\S\ref{sec:sovereignty}.

\subsubsection{Optical Transport Network}

Sites are interconnected by an optical transport network represented
as a directed graph $\mathcal{G} = (\mathcal{S}, \mathcal{E})$,
where each node $s_i \in \mathcal{S}$ corresponds to a data center
site and each directed edge $(s_i, s_j) \in \mathcal{E}$ represents
a provisioned optical path from site $i$ to site $j$.
Directionality reflects the physical reality that each transmission
direction occupies a distinct set of wavelength channels with
independently provisioned capacity. Each edge is characterized by
two parameters:

\begin{itemize}
  \item $C_{ij}$: the transmission capacity \textnormal{[Gbps]},
    determined by the spectral efficiency of the deployed
    transceivers and the number of provisioned wavelengths on the
    optical path;
  \item $d_{ij}$: the one-way propagation delay \textnormal{[ms]},
    equal to the physical fiber route length divided by the speed
    of light in the medium ($c/n$, where $n \approx 1.468$ for
    standard single-mode fiber at 1550\,nm, consistent with
    ITU-T\,G.652~\cite{ITUTG652}).
\end{itemize}

The end-to-end latency for a workload routed along a multi-hop path
$\pi = (s_{i_1}, s_{i_2}, \dots, s_{i_k})$ is the sum of
propagation delays across consecutive hops:
\begin{equation}
  L(\pi) = \sum_{\ell=1}^{k-1} d_{i_\ell,\, i_{\ell+1}}.
  \label{eq:latency}
\end{equation}
Throughout this paper, $\lambda_k$ denotes a \emph{one-way}
propagation budget, so the latency admissibility condition
$L(\pi) \leq \lambda_k$ is evaluated against the one-way
sum~\eqref{eq:latency}. This convention is applied consistently across all workload classes; operators adopting a round-trip convention must halve their Service Level Objective (SLO) values before applying the
constraints in \S\ref{subsec:opt_problem} and the geographic radius expressions in \S\ref{subsec:fsor_definition}.

This formulation treats propagation delay as the dominant latency
component, which is justified for geographically distributed
placements where inter-site distances are of order
$10^2$--$10^3$\,km. At these scales, processing and queuing delays
are negligible relative to propagation and are omitted from the
placement model (this assumption breaks down for intra-metro or
intra-campus deployments, where propagation delays fall below
1\,ms and switching and queuing overhead become the dominant
contributors). Where empirical hardware characterization is
available, processing and queuing contributions may be incorporated
as site-level additive offsets to $d_{ij}$ without altering the
structure of the formulation.

Propagation delay $d_{ij}$ constitutes a hard physical floor: no
routing algorithm, protocol optimization, or hardware improvement
can reduce end-to-end latency below the speed-of-light bound
imposed by the fiber route geometry~\cite{singla2014}. This
irreducibility has a structural consequence for the FSOR. The
latency constraint $L(\pi) \leq \lambda_k$ partitions the set of
candidate placement--routing pairs into a latency-admissible subset
whose boundary is determined entirely by the physical topology of
the optical network and the geographic distribution of demand---not
by the sustainability profile of individual sites. A site may be
fully compliant with all carbon and water constraints and yet be
unreachable for a latency-sensitive workload simply because no
fiber path of sufficient brevity connects it to the demand source.
This decoupling between sustainability eligibility and latency
admissibility is the geometric origin of the green-but-far effect
discussed in \S\ref{subsec:endowments}.

\subsubsection{Workload Characterization}

Let $\mathcal{W} = \{w_1, \dots, w_M\}$ denote the set of $M$ active
workloads. Each workload $w_k$ is characterized by:
\begin{itemize}
  \item $p_k$: the power demand \textnormal{[kW]};
  \item $\lambda_k$: the maximum tolerable end-to-end latency
    \textnormal{[ms]};
  \item $\rho_k$: the traffic demand \textnormal{[Gbps]} injected onto
    the network;
  \item $\mu_k \in \{0,1\}$: a portability indicator, equal to $1$ if the
    workload can be spatially shifted without violating synchronization or
    state-transfer constraints, and $0$ otherwise.
\end{itemize}

In practice, portability is not a clean binary. It depends on latency tolerance, inter-replica synchronization requirements, and the overhead of transferring workload state across sites. For LLM workloads, the “state” includes large data that must move with the workload. This includes model files, which are the trained weights of the model, checkpoints, which are intermediate saved states used to resume training or inference, and dataset shards, which are partitions of large datasets distributed across storage. Because these components can total hundreds of gigabytes or more, migrating an LLM workload requires transferring substantial data. This makes migration not only time-sensitive, but also cost- and carbon-sensitive due to network usage and energy consumption~\cite{green_scheduling_llm,water_efficient_llm}. We treat $\mu_k$ as a conservative binary indicator---a workload is
portable only if its migration overhead can be bounded within acceptable
latency and carbon budgets---and note that a continuous portability score
could be substituted without altering the structure of the placement
problem. The binary treatment is retained throughout the remainder of
this paper; the continuous generalization is identified as a direction
for future work in~\S\ref{subsec:limitations}.

Workload classes differ systematically in their parameter distributions
and constraint profiles:

\begin{description}
  \item[Training workloads] are characterized by high $p_k$, low
    $\mu_k$, and high inter-accelerator bandwidth requirements that
    impose tight constraints on both intra-site fabric and inter-site
    routing. They are spatially constrained by the need for low-latency
    collective communication (AllReduce, AllGather) across accelerator
    ranks~\cite{wei2025distributed}. Once placed, they are effectively
    immobile for the duration of the training run.

  \item[Inference workloads] have moderate $p_k$ and moderate
    portability in principle, but are latency-constrained in practice:
    the strict $\lambda_k$ imposed by user-facing service-level
    objectives limits the geographic range of admissible placements to
    sites reachable within the one-way propagation budget. Violations
    are directly observable and contractually
    relevant~\cite{gujarati2020}.

  \item[Batch analytics workloads] are delay-tolerant ($\lambda_k$
    large or unconstrained), highly portable ($\mu_k = 1$), and
    represent the primary candidates for temporal and spatial shifting
    in response to carbon intensity or water stress signals. They form
    the most responsive layer of the workload portfolio for
    sustainability-driven scheduling~\cite{Sukprasert2024}.
\end{description}

\subsubsection{Migration Overhead}

Moving a workload from one data center to another is not
instantaneous, even when the destination site is
sustainability-compliant and the network path is available.
Migration consumes time, energy, and network capacity---and a
relocation that appears attractive on sustainability grounds may
be inadvisable if its overhead erodes the very gains it was
intended to achieve.\footnote{The analogy to relocating a running
business is instructive: filing cabinets (model weights and
execution state) must be transported, the move itself consumes
fuel (grid energy at the source), and staff cannot resume full
productivity until systems are reinstalled and institutional memory
is reconstructed (cache warming, session re-establishment). Each
step takes time and consumes resources; the business is not fully
operational until all three are complete.}

For workloads with $\mu_k = 1$, cross-site migration from $s_i$
to $s_j$ incurs overheads beyond the steady-state propagation
delay in~\eqref{eq:latency}. We model three components:

\begin{enumerate}
  \item \textbf{Data transfer delay}: proportional to the
    workload state size $\sigma_k$~\textnormal{[GB]} and
    inversely proportional to the available transfer bandwidth
    $B_{ij}(t)$~\textnormal{[Gbps]}:
    \begin{equation*}
      \delta^{\mathrm{tx}}_{ijk}(t)
        = \frac{8\,\sigma_k}{B_{ij}(t)}
        \quad \textnormal{[s]},
    \end{equation*}
    where the factor of 8 converts $\sigma_k$ from gigabytes to
    gigabits so that numerator and denominator share the same
    unit \textnormal{[Gb]} and \textnormal{[Gbps]} respectively,
    yielding a delay in seconds.

  \item \textbf{Transfer energy}: the electrical energy consumed
    transmitting $\sigma_k$ across the optical link contributes a
    carbon cost
    \begin{equation*}
      \Delta C^{\mathrm{tx}}_{ijk}(t)
        = \gamma_i(t)\cdot e_{ij}\cdot 8\,\sigma_k
        \quad \textnormal{[gCO$_2$eq]},
    \end{equation*}
    where $e_{ij}$~\textnormal{[J/bit]} is the link energy per
    bit and $8\,\sigma_k$~\textnormal{[Gb]} is the data volume
    transmitted. The carbon intensity $\gamma_i(t)$ of the
    \emph{source} site $s_i$ is used here because the
    transmission energy is drawn from the source-side grid
    during the transfer window; energy consumed at intermediate
    optical amplifiers is absorbed into $e_{ij}$ as a per-bit
    cost and does not require a separate intensity term.

  \item \textbf{Rehydration latency}: the time required to
    restore execution context at the destination site, which is
    workload-specific and may include model loading, cache
    warming, and session re-establishment. Rehydration latency
    is not captured by $\delta^{\mathrm{tx}}_{ijk}$ and must be
    accounted for separately when tightening the effective
    latency budget at the destination; see
    constraint~\eqref{eq:latency_migration}.
\end{enumerate}

These overheads are especially consequential for stateful (workloads that maintain and depend on internal state such as model parameters, session data, or intermediate computations that must be preserved across execution and migration) or model-intensive workloads~\cite{water_efficient_llm,green_scheduling_llm}. In the placement optimization, migration costs enter as additive penalties in the objective function and as tightened effective latency budgets: a workload with $\lambda_k = 50$\,ms that
requires $\delta^{\mathrm{tx}}_{ijk} = 30$\,ms to migrate
retains only 20\,ms of latency headroom at the destination before
its service-level objective is violated.

\subsection{Optimization Problem}
\label{subsec:opt_problem}

The framework makes two classes of decisions simultaneously: which site each workload should be placed at, and how the traffic generated by each workload should be routed through the optical network. These decisions are coupled: the carbon and water impact of a placement depends on which site is selected, while the latency experienced by a workload depends on the routing path, which in turn depends on the placement. Neither decision can be made optimally in isolation.

\subsubsection{Decision Variables}
The optimization makes two coupled decisions at each telemetry cycle: \emph{where} to run each workload, and \emph{how} to route the traffic it generates through the optical network. These are represented by two
distinct variable classes, reflecting the fundamental difference between a discrete placement decision --- a workload class is assigned to exactly one sovereign site selected from the feasible sovereign operating region --- and a continuous flow decision, where traffic can
be split across multiple optical paths in arbitrary proportions.

Let $x_{ik} \in \{0,1\}$ denote the placement variable, equal to $1$ if workload $w_k$ is assigned to site $s_i$ and $0$ otherwise. The domain of $x_{ik}$ is restricted to FSOR-eligible sites: only sites $s_i \in \mathcal{S}_{\mathrm{FSOR}}$ satisfying the carbon, water, and power feasibility conditions are admitted as
candidates. For a given workload $w_k$, the vector
$(x_{1k}, x_{2k}, \dots, x_{Nk})$ encodes which of the $N$ candidate sites hosts it; the unique-assignment
constraint~\eqref{eq:assignment} ensures that exactly one entry equals $1$. The binary nature of $x_{ik}$ reflects the operational reality that a workload has a single authoritative placement at any given time,
and the FSOR restriction ensures that every feasible solution is sovereign by construction --- no post-hoc feasibility filter is required.

Let $f_{ijk}(t) \geq 0$ denote the continuous flow variable
representing the traffic volume~\textnormal{[Gbps]} of workload $w_k$ routed along directed edge $(s_i, s_j)$ at telemetry cycle $t$. Unlike placement, routing admits fractional solutions: traffic from a single workload may be split across multiple optical paths, and the flow
variables capture this flexibility. The time index $t$ reflects that routing decisions are re-evaluated at each telemetry cycle in response to changing link utilization and workload demand.

The full decision vector is $(\mathbf{x}, \mathbf{f})$, where $\mathbf{x} \in \{0,1\}^{N \times M}$ collects all placement decisions across the $N$ FSOR-eligible sites and $M$ workloads, and $\mathbf{f} \in \mathbb{R}_{\geq 0}^{|\mathcal{E}| \times M}$ collects
all routing flows across the $|\mathcal{E}|$ directed edges and $M$ workloads. The two components are not independent: the placement $\mathbf{x}$ determines the source node of each workload's traffic, which in turn constrains the feasible flow patterns in $\mathbf{f}$ through the flow-conservation constraint~\eqref{eq:flow_conservation}. This coupling is the formal reason why placement and routing must be optimized jointly rather than sequentially.
\subsubsection{Objective Function}

The objective is to minimize the aggregate environmental impact
across all active workloads, combining carbon emissions and water
consumption into a single weighted criterion:
\begin{equation}
  \min_{\mathbf{x},\,\mathbf{f}} \;\;
  \alpha \sum_{i \in \mathcal{S}} \sum_{k \in \mathcal{W}}
    \tilde{\gamma}_i(t)\, p_k\, x_{ik}
  \;+\;
  (1-\alpha) \sum_{i \in \mathcal{S}} \sum_{k \in \mathcal{W}}
    \tilde{\omega}_i(t)\, p_k\, x_{ik},
  \label{eq:objective}
\end{equation}
where $\tilde{\gamma}_i(t)$ and $\tilde{\omega}_i(t)$ are the
site-level carbon intensity and water usage effectiveness,
normalized to a common dimensionless scale so that the weighted
sum is commensurable.\footnote{Normalization is performed
site-wise using min-max scaling across the observed range of
$\gamma_i$ and $\omega_i$ over the planning horizon:
$\tilde{\gamma}_i = (\gamma_i - \gamma_{\min}) /
(\gamma_{\max} - \gamma_{\min})$, and analogously for
$\tilde{\omega}_i$. This maps each parameter to $[0,1]$
independently, eliminating the unit incompatibility between
\textnormal{[gCO$_2$eq/kWh]} and \textnormal{[L/kWh]} without
requiring a physical conversion factor. The companion
optimization paper provides a sensitivity analysis of objective
sensitivity to the choice of normalization method.}
In the non-normalized parameterization,
$\gamma_i(t)$~\textnormal{[gCO$_2$eq/kWh]} and
$\omega_i(t)$~\textnormal{[L/kWh]} are obtained directly from
grid telemetry and facility monitoring respectively, and
$p_k$~\textnormal{[kW]} is the power demand of workload $w_k$.
The scalar $\alpha \in [0,1]$ is an operator-defined weight
reflecting the relative priority of carbon versus water
reduction; it absorbs any residual unit-scaling factor required
to make the two terms comparable. Setting $\alpha = 1$ recovers
a pure carbon minimization objective; $\alpha = 0$ minimizes
water consumption alone; intermediate values allow
Pareto-efficient trade-offs between the two dimensions to be
traced by parametric variation of $\alpha$.

The routing variables $\mathbf{f}$ do not appear explicitly in
the objective. \textit{Assumption~A1}: network transport energy
is treated as a second-order contributor to site-level carbon
cost relative to compute load at the inter-site distances
considered. This assumption is motivated by the energy profile
of modern long-haul optical transport, where amplified WDM
links operate at link-level energy intensities of order
$10^{-1}$--$10^{0}$~pJ/bit~\cite{Tucker2011power}, several
orders of magnitude smaller than the per-bit energy equivalent
of GPU compute at typical AI workload power densities. The
assumption may not hold at very high migration data volumes
(large $\sigma_k$) or on legacy links with high energy per bit;
in those configurations, a network transport term of the form
$\sum_{(i,j)\in\mathcal{E}} \sum_{k\in\mathcal{W}}
\gamma_i(t)\, e_{ij}\, f_{ijk}(t)$ can be added
to~\eqref{eq:objective} without altering the constraint
structure. Under Assumption~A1, routing decisions are governed
entirely by the capacity and latency
constraints~\eqref{eq:link_capacity}--\eqref{eq:flow_conservation}.

The formulation focuses on carbon and water as the primary
environmental objectives, reflecting that these two dimensions
are subject to the hardest regulatory constraints in current and
emerging frameworks and are most directly tied to physically
observable telemetry parameters. Geo-distributed scheduling
frameworks that additionally incorporate energy cost and
quality-of-service metrics confirm the inherently
multi-dimensional nature of sustainable AI orchestration and
represent natural extensions of the present
formulation~\cite{hoxha2025llm}.

\subsubsection{Hard Constraints versus Penalty Encoding}
\label{subsec:hard_vs_penalty}

The central modeling decision is the treatment of sustainability
limits. Rather than penalizing high-carbon or high-water
placements in the objective function---a soft encoding that allows
violations at sufficient cost---the framework encodes them as
\emph{hard constraints} that eliminate non-compliant
configurations from the feasible region entirely. This distinction
is not merely technical: it reflects the legal and physical
reality that regulatory maxima are not preferences to be traded
against, but absolute limits whose violation carries legal,
financial, and reputational consequences that cannot be offset by
environmental performance on other dimensions.

The carbon constraint eliminates site $s_i$ from consideration
for all workloads whenever its grid intensity exceeds the
site-level threshold. In the MILP, this is implemented as a
preprocessing variable fixing evaluated from the telemetry
snapshot before the solve:
\begin{equation}
  x_{ik} \;\leq\; \mathbf{1}\bigl[\gamma_i(t) \leq
  \bar{\Gamma}_i\bigr]
  \qquad \forall\, i \in \mathcal{S},\; k \in \mathcal{W},
  \label{eq:carbon_gate}
\end{equation}
where $\mathbf{1}[\cdot]$ is an indicator that equals $1$ if the condition holds and $0$ otherwise, and is evaluated from telemetry data prior to invoking the solver. When $\gamma_i(t) > \bar{\Gamma}_i$, the upper bound on $x_{ik}$ is set to zero for all $k$, effectively removing site $s_i$ from the feasible set before optimization begins. Commercial solvers such as Gurobi and CPLEX handle such bound fixings in the presolve phase at negligible computational cost, and the reduced problem passed to the branch-and-bound engine is smaller by all variables associated with the excluded site~\cite{Achterberg2013}. The carbon gate is site-level: once $\gamma_i(t)$ exceeds $\bar{\Gamma}_i$, the site is ineligible for any workload regardless of power demand, because grid carbon intensity is independent of what is being
computed.

Water draw is proportional to the aggregate IT load placed at a site, not to individual workload assignments in isolation.
The correct formulation is therefore an aggregate linear
inequality, analogous to the power capacity
constraint~\eqref{eq:power}:
\begin{equation}
  \omega_i(t) \sum_{k \in \mathcal{W}} p_k\, x_{ik}
  \;\leq\; \bar{\Omega}_i(t)
  \qquad \forall\, i \in \mathcal{S}.
  \label{eq:water_capacity}
\end{equation}
This constraint ensures that the total water draw at site $s_i$
--- the product of water usage effectiveness $\omega_i(t)$
\textnormal{[L/kWh]} and aggregate IT power
\textnormal{[kW]} --- does not exceed the site's permit limit
$\bar{\Omega}_i(t)$ \textnormal{[L/h]}. Unlike the carbon gate,
which produces a binary site exclusion, \eqref{eq:water_capacity}
defines a polyhedral feasibility boundary over the placement vector: as the aggregate power of co-located workloads approaches the permit headroom, additional placements at that site become infeasible, but placements that individually respect the remaining headroom remain admissible. This is qualitatively distinct from the all-or-nothing carbon exclusion: a site subject to a binding water constraint is not eliminated from the feasible set but rather has its effective hosting capacity reduced as a function of the workload composition assigned to it.

The structural difference between \eqref{eq:carbon_gate} and \eqref{eq:water_capacity} has a direct consequence for the FSOR geometry. Carbon thresholds produce discrete, discontinuous contractions of the feasible set: a site is either eligible or it is not, and a marginal increase in $\gamma_i(t)$ past $\bar{\Gamma}_i$ removes it
entirely. Water constraints, by contrast, produce a progressive polyhedral tightening that depends on the composition of the active workload set: adding a power-intensive workload to a site may exhaust the remaining permit headroom and render subsequent placements at that site infeasible, even though no individual workload alone exceeds the limit.

To see why, consider that the water constraint accumulates across all workloads assigned to site $s_i$: each placement $x_{ik} = 1$ consumes a share $\omega_i(t)\, p_k$
of the site's water permit $\bar{\Omega}_i(t)$, and the residual
headroom shrinks accordingly. A workload $w_{k''}$ that would be
perfectly admissible in isolation --- because
$\omega_i(t)\, p_{k''} \leq \bar{\Omega}_i(t)$ --- may become
infeasible once a prior power-intensive assignment $x_{ik'} = 1$
has already claimed a large fraction of that budget. In geometric
terms, each placement tightens one face of the feasible polytope in
$\mathbf{x}$-space, progressively shrinking the region of admissible
joint assignments. This is qualitatively different from the carbon
gate, which acts as a site-level switch independent of workload
composition: no combination of workloads at site $s_i$ can
rehabilitate a site whose grid intensity exceeds $\bar{\Gamma}_i$.
The practical implication is that water feasibility is an emergent
property of the \emph{joint} placement decision $\mathbf{x}$, not a
property that can be checked workload by workload in isolation.

These constraints produce the discrete and polyhedral FSOR
geometry described in \S\ref{subsec:fsor_metric}. The
soft-penalty alternative---adding
$\beta_\gamma \max(0,\, \gamma_i(t) - \bar{\Gamma}_i)$ and
$\beta_\omega \max\!\bigl(0,\, \omega_i(t)\sum_k p_k x_{ik}
- \bar{\Omega}_i(t)\bigr)$ to the objective---preserves the
convex structure of the relaxed problem and may be preferable in settings where constraints represent operational targets
rather than regulatory floors, or where the hard feasible set would otherwise be empty and a best-effort solution is required. However, the penalty weights $\beta_\gamma$ and $\beta_\omega$ require calibration that is neither transparent to the operator nor derivable from physical or regulatory parameters: their choice determines whether violations are tolerated in practice, and a poorly chosen $\beta$ can produce solutions that nominally minimize the objective while systematically violating regulatory limits at low penalty cost. The choice between hard and soft encoding is therefore a design parameter that must be made explicit and justified against the regulatory context of the deployment. We adopt hard constraints throughout as the conservative default, consistent with the interpretation of carbon and water limits as binding regulatory obligations rather than cost terms.

\subsubsection{Constraint Set}

The notation $o(k)$ and $d(k)$, used throughout the routing
constraints below, denote respectively the \emph{source index} of workload $w_k$---the site at which it is placed, as determined by $\mathbf{x}$---and the \emph{destination index} of the demand-serving endpoint to which its traffic must be delivered. These indices are workload-specific functions of the placement vector and are used consistently in the flow conservation constraint~\eqref{eq:flow_conservation} and in the FSOR definition of \S\ref{subsec:fsor_definition}.

Subject to the sustainability
gate~\eqref{eq:carbon_gate} and the water capacity
constraint~\eqref{eq:water_capacity}, the optimization must
simultaneously satisfy the following constraints.

Unique assignment: each workload is placed at exactly one site:
\begin{equation}
  \sum_{i \in \mathcal{S}} x_{ik} = 1
  \qquad \forall\, k \in \mathcal{W}.
  \label{eq:assignment}
\end{equation}

Power capacity: the aggregate power demand of workloads placed at site $s_i$
must not exceed its available electrical capacity:
\begin{equation}
  \sum_{k \in \mathcal{W}} p_k\, x_{ik} \leq P_i(t)
  \qquad \forall\, i \in \mathcal{S}.
  \label{eq:power}
\end{equation}

Latency: the end-to-end propagation latency experienced by workload
$w_k$ must not exceed its service-level objective $\lambda_k$:
\begin{equation}
  L(\pi_k) \leq \lambda_k
  \qquad \forall\, k \in \mathcal{W},
  \label{eq:latency_slo}
\end{equation}
where $\pi_k$ is the routing path selected for $w_k$ and
$L(\pi_k)$ is defined by~\eqref{eq:latency}. In the MILP, this
constraint is linearized using an arc-based formulation in which
the latency accumulated along active flow arcs is tracked as an
auxiliary variable, following standard practice for
delay-constrained flow problems~\cite{Gouveia1996}.

For portable workloads ($\mu_k = 1$) subject to migration, the
effective latency budget is reduced by both the data transfer
delay $\delta^{\mathrm{tx}}_{ijk}$ and the rehydration latency
$\delta^{\mathrm{rh}}_{jk}$ introduced in \S\ref{subsec:infra_model}.
The admissible placement set at the destination is therefore
restricted to sites satisfying:
\begin{equation}
  L(\pi_k)
  + \delta^{\mathrm{tx}}_{ijk}(t)
  + \delta^{\mathrm{rh}}_{jk}
  \leq \lambda_k.
  \label{eq:latency_migration}
\end{equation}
Here $\delta^{\mathrm{rh}}_{jk}$~\textnormal{[s]} denotes the
rehydration latency at destination site $s_j$ for workload
$w_k$, covering model loading, cache warming, and session
re-establishment. Both overhead components are subtracted from
the available latency budget before the routing constraint is
evaluated; a workload with $\lambda_k = 50$\,ms that requires
$\delta^{\mathrm{tx}}_{ijk} = 20$\,ms and
$\delta^{\mathrm{rh}}_{jk} = 10$\,ms to migrate retains only
20\,ms of propagation headroom at the destination.

Network capacity: the aggregate traffic on each directed optical link must not
exceed its transmission capacity:
\begin{equation}
  \sum_{k \in \mathcal{W}} f_{ijk}(t) \leq C_{ij}
  \qquad \forall\, (s_i, s_j) \in \mathcal{E}.
  \label{eq:link_capacity}
\end{equation}

Flow conservation: routing flows must be conserved at every node. Using the
net-outflow convention, the flow balance at each site $s_i$
for workload $w_k$ at telemetry cycle $t$ is:
\begin{equation}
  \sum_{j:\,(s_i,s_j)\in\mathcal{E}} f_{ijk}(t)
  \;-\;
  \sum_{j:\,(s_j,s_i)\in\mathcal{E}} f_{jik}(t)
  \;=\;
  \begin{cases}
    +\rho_k & \text{if } i = o(k), \\
    -\rho_k & \text{if } i = d(k), \\
    \phantom{+}0 & \text{otherwise,}
  \end{cases}
  \label{eq:flow_conservation}
\end{equation}
where $\rho_k$~\textnormal{[Gbps]} is the traffic demand of
workload $w_k$. The left-hand side counts net flow leaving site
$s_i$ for workload $w_k$: it is positive at the source
(traffic is injected), negative at the destination (traffic is
absorbed), and zero at all intermediate transit nodes. Routing
decisions are re-evaluated at each telemetry cycle $t$; the
time index is carried consistently on all flow variables
$f_{ijk}(t)$ throughout this formulation.

\begin{remark}[Source index linearization]
The source index $o(k)$ in~\eqref{eq:flow_conservation} depends
on the placement vector $\mathbf{x}$, which is itself a decision
variable. Naively substituting $o(k)$ as a function of
$\mathbf{x}$ would introduce a bilinear product between the
binary placement variables and the continuous flow variables,
destroying the linear structure of the MILP. This is resolved by
expanding the right-hand side of~\eqref{eq:flow_conservation}
using a standard big-$M$ linking formulation: for each candidate
source site $s_i$, the injection term $+\rho_k$ is activated if
and only if $x_{ik} = 1$, yielding the equivalent linear
constraint
\begin{equation}
  \sum_{j:\,(s_i,s_j)\in\mathcal{E}} f_{ijk}(t)
  \;-\;
  \sum_{j:\,(s_j,s_i)\in\mathcal{E}} f_{jik}(t)
  \;=\;
  \rho_k \bigl(x_{ik} - x_{d(k),k}\bigr)
  \qquad \forall\, i \in \mathcal{S},\; k \in \mathcal{W},
  \label{eq:flow_conservation_linear}
\end{equation}
where $x_{d(k),k} = 1$ if site $s_i$ is the destination of
workload $w_k$ and $0$ otherwise. Because $x_{ik} \in \{0,1\}$,
the right-hand side evaluates to $+\rho_k$ at the unique source,
$-\rho_k$ at the destination, and $0$ at all other sites,
recovering the case structure of~\eqref{eq:flow_conservation}
exactly. The linearized form~\eqref{eq:flow_conservation_linear}
is what is passed to the solver; \eqref{eq:flow_conservation} is
retained in the text for expository clarity.
\end{remark}
\subsubsection{The Green-But-Far Tension}
\label{subsec:tension}

The interaction between the sustainability
gate~\eqref{eq:carbon_gate} and the water capacity
constraint~\eqref{eq:water_capacity} on one side, and the
latency constraint~\eqref{eq:latency_slo} on the other, is the central tension the framework must navigate, and gives rise to what we term the \emph{green-but-far effect}: the sites that are most sustainability-compliant are often geographically remote from demand sources, and the propagation delay floor imposed by the fiber route geometry may place them beyond the latency budget of sensitive workloads. The effect operates in both
directions. Excluding high-carbon or high-water sites from the feasible set may force workloads onto geographically distant sites with longer optical paths, potentially violating latency service-level objectives. Conversely, anchoring latency-sensitive workloads to nearby sites may leave no local option that satisfies carbon or water thresholds, producing a feasibility gap that neither placement nor routing optimization alone can close. The sovereignty implications of this effect --- in particular, the structural asymmetry it creates between latency-sensitive and delay-tolerant workload classes --- are
developed in~\S\ref{subsec:sovereignty_mechanisms}.

\begin{figure}[t]
  \centering
  \includegraphics[width=\linewidth]{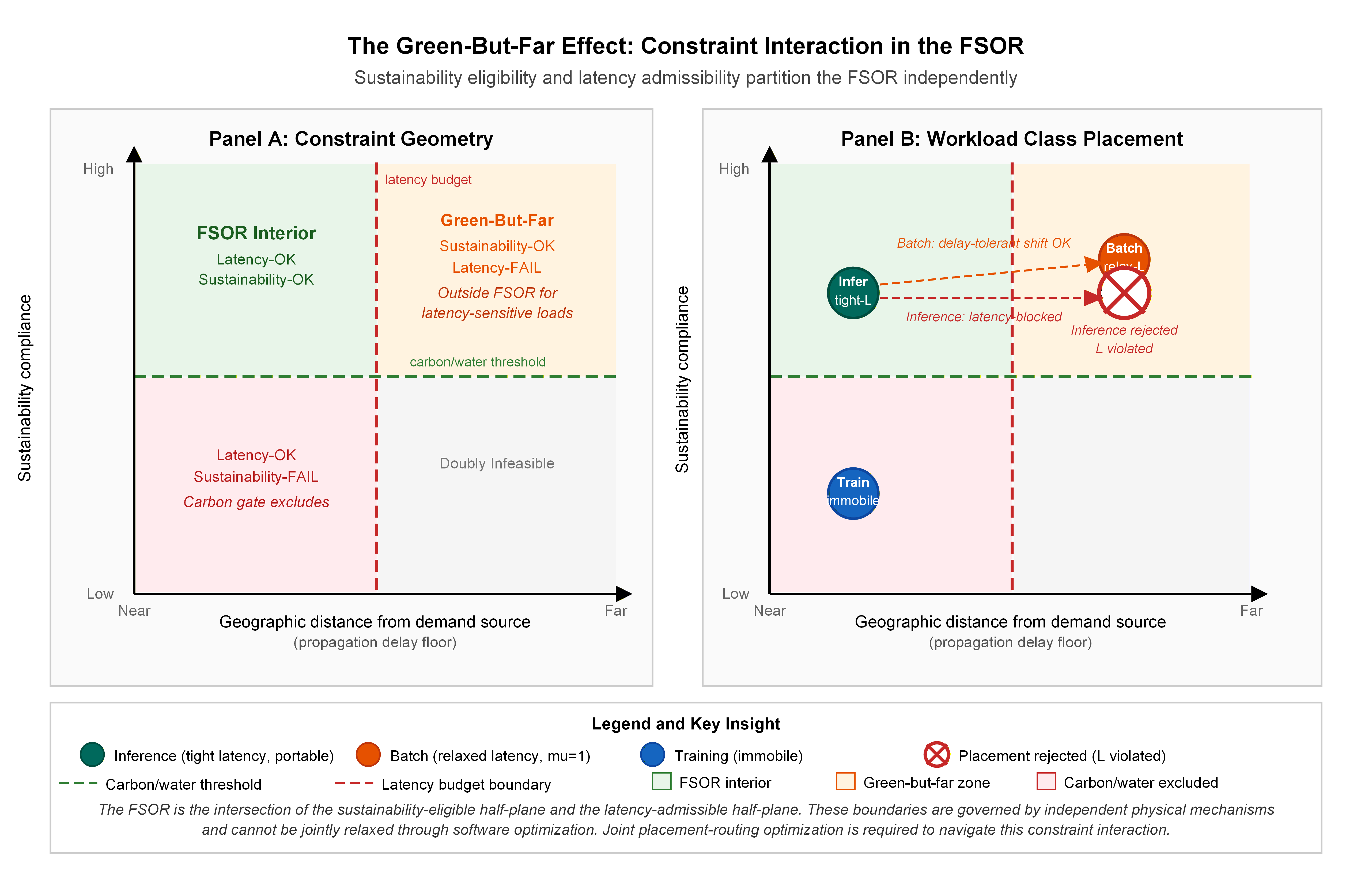}
  \caption{The green-but-far effect as a constraint geometry.
    \textbf{Panel~A} shows the two-dimensional space of candidate
    sites partitioned by the carbon/water sustainability threshold
    $\bar{\Gamma}_i$ (green dashed line) and the latency budget
    $\lambda_k$ (red dashed line). The FSOR interior (green
    region) is the intersection of the sustainability-eligible and latency-admissible half-planes; the green-but-far zone (orange region) contains sites that satisfy all sustainability constraints but lie beyond the propagation-delay radius admitted by the latency SLO. \textbf{Panel~B} maps the three workload classes onto the same constraint space. Inference workloads (tight latency SLO) are confined to the FSOR interior and cannot reach green-but-far sites regardless of their sustainability profile. Batch workloads (relaxed latency) can exploit
    delay-tolerant shifting to reach sustainability-compliant remote sites. Training workloads are immobile once placed and cannot be shifted in response to sustainability signals. The two boundary lines are governed by independent physical mechanisms --- grid carbon intensity and fiber propagation geometry --- and cannot be jointly relaxed through software optimization alone; this is the structural reason why joint placement--routing optimization is required.}
  \label{fig:green_but_far}
\end{figure}

This tension cannot be resolved by optimizing placement and
routing independently. A placement that appears optimal in
isolation may induce routing paths that violate latency or link
capacity constraints; a routing that is efficient given a fixed
placement may become infeasible when that placement is revised
under sustainability pressure. Joint optimization over
$(\mathbf{x}, \mathbf{f})$ is therefore not merely a
computational convenience but a structural necessity: it is the
minimal formulation in which the green-but-far effect is
representable as a constraint interaction rather than a post-hoc
correction. The coupling is encoded in
constraint~\eqref{eq:latency_slo}, which links the binary
placement variable $x_{ik}$ to the continuous routing path
$\pi_k$ through the propagation delay along the selected path,
making the two decisions jointly constrained rather than
separable.

When no feasible $(\mathbf{x}, \mathbf{f})$ exists that
satisfies all constraints simultaneously, the current workload
configuration lies outside the FSOR under the prevailing
telemetry snapshot. This is not an error condition but an
informative signal: it identifies the precise combination of
constraints whose simultaneous binding renders operation
infeasible, providing the operator with a minimal description of
what would need to change to restore feasibility. In operational
deployments, infeasibility should trigger either workload
deferral (for delay-tolerant batch workloads, which can wait for
a more favorable telemetry snapshot) or an explicit constraint
relaxation protocol in which the operator selects which limit to
relax and by how much, guided by the binding constraint set
returned by the solver. The formal structure of this
relaxation---and its interpretation as an Irreducible Infeasible
Set---is discussed in~\S\ref{subsec:hard_vs_penalty_disc}.
Infeasibility is treated as part of the solver output rather
than pre-empted through objective penalization, consistent with
the hard-constraint encoding adopted throughout.

\subsubsection{Problem Class and Tractability}
\label{subsubsec:tractability}

The formulation~\eqref{eq:objective}--\eqref{eq:flow_conservation} is a \emph{mixed-integer linear program} (MILP)~\cite{Bixby2012}: the placement variables $x_{ik}$ are binary, the routing flow variables $f_{ijk}(t)$ are continuous, and all constraints are linear in $(\mathbf{x}, \mathbf{f})$.
The integrality of $\mathbf{x}$ makes the problem NP-hard in general, by reduction from capacitated facility
location~\cite{GareyJohnson1979,Cornuejols1990}: the workload placement subproblem shares the same binary assignment structure as the capacitated plant location problem, and NP-hardness follows directly from that reduction. The block structure---binary placement coupled to continuous routing through the capacity and latency constraints---is, however, well suited to decomposition. Tractability at these scales is driven by the strong pruning induced by the sustainability and latency gates, which reduce the effective binary search space before branch-and-bound exploration.

For the scenario sizes considered in this work (up to $N = 8$ sites and $|\mathcal{W}| = 20$ workload classes per telemetry cycle), the problem is tractable with open-source MILP solvers within operationally relevant timescales. In Python/PuLP experiments on a standard workstation, all representative instances at these dimensions were solved to certified optimality
(Table~\ref{tab:benchmark}): median solve time was under
0.2\,s for $N \leq 6$, and under 60\,s for the full $N=8$,
$|\mathcal{W}|=20$ scale---well within a five-minute telemetry cycle. Scaling to continental deployments with $N \gg 10^2$ candidate sites, fine-grained routing topologies, and rolling time horizons requires more scalable solution methods.

\begin{table}[t]
\centering
\caption{Solve-time benchmark for the FSOR MILP
         (\S\ref{subsec:opt_problem}), solved with the open-source
         CBC solver via Python/PuLP~3.3 on a standard workstation.
         Each row reports statistics over five randomised instances
         drawn from the same parameter distributions used in
         \S\ref{sec:evaluation}. All instances were solved to
         certified optimality. The telemetry cycle budget is
         300\,s (5\,min).}
\label{tab:benchmark}
\small
\begin{tabular}{lccrrcrrrr}
\toprule
& & &
  \multicolumn{2}{c}{Variables} & &
  \multicolumn{4}{c}{Solve time [s]} \\
\cmidrule{4-5} \cmidrule{7-10}
Scenario & $N$ & $M$ &
  Binary$^{a}$ & Continuous$^{b}$ &
  Opt.$^{c}$ &
  Min & Median & Mean & Max \\
\midrule
Small  & 4 &  10 & $\leq 40$  &     120 & 5/5 &
  0.02 & 0.02 &  0.03 & 0.06 \\
Medium & 6 &  15 & $\leq 90$  &     450 & 5/5 &
  0.04 & 0.07 &  0.17 & 0.58 \\
Paper  & 8 &  20 & $\leq 140$ & 1\,120 & 5/5 &
  0.11 & 0.12 & 18.39 & 60.00 \\
\midrule
\multicolumn{6}{l}{Telemetry cycle budget (5\,min)} &
  \multicolumn{4}{r}{300\,s} \\
\bottomrule
\end{tabular}

\medskip
\begin{minipage}{0.94\linewidth}\footnotesize
$^{a}$~Active binary placement variables $x_{ik}$ after
preprocessing: the carbon gate~\eqref{eq:carbon_gate} and
latency gate~\eqref{eq:latency_slo} fix infeasible $x_{ik}=0$
before the solver is invoked, reducing the binary search space
below the nominal $N \times M$ count.\\[2pt]
$^{b}$~Continuous routing flow variables $f_{ijk}(t)$; equal to
$|\mathcal{E}| \times M$ for the complete directed graph
($|\mathcal{E}| = N(N-1)$) used here.\\[2pt]
$^{c}$~Trials solved to certified optimality out of five; all
five instances at every scale were solved to optimality. The two
longest runs in the $N=8$, $M=20$ scenario (31\,s and 60\,s)
reflect worst-case branching behavior on specific random seeds;
both remain well within the 300\,s telemetry cycle budget.
\end{minipage}
\end{table}

Benders decomposition provides a natural way to exploit the structure of the problem by separating placement and routing decisions into two interacting subproblems~\cite{Benders1962,Rahmaniani2017}. The method proceeds iteratively. First, a \emph{master problem} selects a candidate placement, assigning workloads to sites subject to sustainability and capacity constraints. Given this placement, a \emph{subproblem} is then solved to determine whether the resulting traffic can be feasibly routed through the optical network under capacity and latency constraints.

If the routing subproblem is feasible, the current placement is valid. If it is infeasible, the subproblem generates a constraint, known as a Benders cut, that rules out the current placement and similar infeasible configurations. This cut is added to the master problem, which is then re-solved to produce a new placement. Through this iterative exchange, the method progressively eliminates infeasible regions of the search space and converges to a jointly feasible solution without solving the full placement–routing problem in a single step~\cite{Bonami2020}.

For settings where strict optimality is not required, Lagrangian relaxation offers a practical way to reduce computational complexity~\cite{Fisher1981}. The key idea is to temporarily relax the constraints that couple placement and routing, and instead incorporate them into the objective function using penalty multipliers. In this relaxed problem, violations of the coupling constraints are allowed, but they incur a cost, which makes the problem easier to solve because it separates into simpler subproblems.

This approach provides a lower bound on the optimal solution, and by adjusting the penalty multipliers iteratively, it is often possible to recover high-quality feasible solutions that are close to optimal. In practice, this means that instead of solving the full joint problem exactly, one can solve a sequence of simpler problems and combine their results.

The effectiveness of Lagrangian relaxation depends on how well the relaxed problem approximates the original one. When the problem structure is well aligned with the relaxation, the gap between the relaxed solution and the true optimum, known as the duality gap, is often small. However, for general mixed-integer problems, especially non-convex ones, this gap cannot be guaranteed to vanish, and solution quality must be evaluated empirically~\cite{Wolsey1998}.

For very large instances where even decomposition-based methods are computationally prohibitive, learning-based approximation offers a complementary route to scalability, at the cost of relinquishing provable feasibility guarantees; this direction is
discussed in~\S\ref{subsec:limitations}.

\subsection{The Feasible Sovereign Operating Region}
\label{subsec:fsor_definition}

\subsubsection{Formal Definition}

Let $\mathcal{X} = \{0,1\}^{N \times M}$ denote the space of
placement configurations and $\mathcal{F} =
\mathbb{R}_{\geq 0}^{|\mathcal{E}| \times M}$ the space of
routing flows. Let $\mathcal{W}_{\mathrm{total}}$ denote the
universe of all workloads that could potentially be submitted
to the infrastructure; the active workload set
$\mathcal{W}(t) \subseteq \mathcal{W}_{\mathrm{total}}$ is the
subset present in the system at telemetry cycle $t$.

Given a telemetry snapshot
\begin{equation*}
  \boldsymbol{\theta}(t) = \bigl\{P_i(t),\, \gamma_i(t),\,
  \omega_i(t),\, \bar{\Omega}_i(t),\, C_{ij},\,
  d_{ij}\bigr\}_{i \in \mathcal{S},\, (i,j) \in \mathcal{E}}
\end{equation*}
that encodes site-level sustainability and power parameters and
network-level capacity and delay parameters, define the
\emph{feasible set} of the constraint
system~\eqref{eq:carbon_gate}--\eqref{eq:flow_conservation} as:
\begin{equation}
  \Phi\bigl(\mathcal{W}(t),\, \boldsymbol{\theta}(t)\bigr)
  \;=\;
  \Bigl\{
    (\mathbf{x}, \mathbf{f}) \in \mathcal{X} \times \mathcal{F}
    \;\Big|\;
    \eqref{eq:carbon_gate}\text{--}\eqref{eq:flow_conservation}
    \text{ are satisfied}
  \Bigr\},
  \label{eq:feasibility_correspondence}
\end{equation}
that is, the set of all placement--routing pairs that
simultaneously satisfy every constraint in
\S\ref{subsec:opt_problem} under the infrastructure state
$\boldsymbol{\theta}(t)$ and active workload set
$\mathcal{W}(t)$. The carbon ceiling $\bar{\Gamma}_i$ is
treated as a fixed regulatory parameter and is not
time-indexed in $\boldsymbol{\theta}(t)$; its evolution over
planning horizons (as regulatory frameworks tighten) is a
distinct multi-period problem discussed in
\S\ref{subsec:fsor_dynamics}. The \emph{Feasible Sovereign
Operating Region} is then defined as the collection of workload
configurations for which this feasible set is non-empty:
\begin{equation}
  \mathrm{FSOR}\bigl(\boldsymbol{\theta}(t)\bigr)
  \;=\;
  \Bigl\{
    \mathcal{W}(t) \subseteq \mathcal{W}_{\mathrm{total}}
    \;\Big|\;
    \Phi\bigl(\mathcal{W}(t),\, \boldsymbol{\theta}(t)\bigr)
    \neq \emptyset
  \Bigr\}.
  \label{eq:fsor_definition}
\end{equation}
The FSOR is a subset of the power set of
$\mathcal{W}_{\mathrm{total}}$, parameterized by the telemetry
state $\boldsymbol{\theta}(t)$. A workload configuration
$\mathcal{W}(t)$ lies inside the FSOR if and only if there
exists at least one valid assignment of workloads to sites and
a corresponding routing of their traffic that jointly satisfies
all physical and regulatory constraints. Crucially, the FSOR
is defined by \emph{joint} feasibility across all constraint
classes---sustainability, latency, power capacity, and network
capacity---not by the satisfaction of any single constraint in
isolation: a configuration that is carbon-compliant but
latency-infeasible lies outside the FSOR just as surely as one
that violates a power cap.

The qualifier \emph{sovereign} reflects an operational
interpretation that goes beyond technical feasibility: the FSOR
characterizes what a region can operate under its own physical
and regulatory endowment without depending on infrastructure or
policy concessions from external actors. The full development of
this interpretation, and its implications for infrastructure
investment and regional AI capacity, is deferred
to~\S\ref{sec:sovereignty}.

\subsubsection{Geometric Structure}

The constraint set of \S\ref{subsec:opt_problem} partitions the
feasible space into four qualitatively distinct regions, whose
geometric character we now describe. The boundary of the FSOR
is not a single smooth surface but the intersection of these
regions, each carved out by a distinct constraint class. Because
the FSOR is defined over the combinatorial space of workload
configurations rather than a continuous vector space, its
``geometry'' is discrete and set-valued: constraints either
admit or exclude entire subsets of configurations, and the
overall feasible set is their intersection.

\begin{description}

  \item[Carbon gate~\eqref{eq:carbon_gate} and water capacity
    constraint~\eqref{eq:water_capacity}.]
    The carbon gate acts as a site-level eligibility filter:
    when $\gamma_i(t)$ exceeds $\bar{\Gamma}_i$, the site is
    removed from the feasible set entirely via the preprocessing
    bound fixing described in \S\ref{subsec:hard_vs_penalty},
    and every workload configuration that requires placement at
    $s_i$---whether because of power, latency, or routing
    considerations---is simultaneously excluded from the FSOR.
    The effect is a discrete, discontinuous contraction of the
    feasible set: the site is either eligible or it is not.
    The water capacity constraint~\eqref{eq:water_capacity}, by
    contrast, defines a linear inequality over the aggregate
    placement vector and produces a polyhedral tightening rather
    than a binary exclusion: as cumulative IT load at a site
    approaches the permit limit, additional placements at that
    site become infeasible while lighter configurations may
    remain admissible.

  \item[Power capacity~\eqref{eq:power}.]
    This constraint defines a set of linear inequalities over
    the placement variables. The feasible region in placement
    space has a polyhedral structure; as power headroom tightens
    at a given site, the feasible configurations involving that
    site are progressively eliminated, while configurations that
    do not rely on it are unaffected. The water capacity and
    power capacity constraints are structurally parallel:
    both impose polyhedral boundaries whose tightness depends
    on the composition of the active workload set assigned to
    each site.

  \item[Latency~\eqref{eq:latency_slo}.]
    This constraint partitions the set of placement--routing
    pairs into latency-admissible and latency-inadmissible
    subsets. This partition is determined entirely by the
    physical topology of the optical network and the geographic
    distribution of demand, independently of the sustainability
    profile of individual sites. Throughout this paper, we
    adopt the one-way propagation convention: $\lambda_k$
    denotes the one-way latency budget, and only sites
    reachable within a propagation delay of $\lambda_k$ are
    admissible for workload $w_k$, corresponding to a geographic
    radius of $\lambda_k \cdot c/n$, where $c/n$ is the speed
    of light in fiber~\cite{singla2014}.

  \item[Network capacity~\eqref{eq:link_capacity} and flow
    conservation~\eqref{eq:flow_conservation}.]
    These constraints impose joint conditions on placement and
    routing that cannot be evaluated on either dimension alone.
    A placement that is individually feasible at each site may
    become globally infeasible when the induced traffic pattern
    exceeds link capacity on shared optical paths, creating
    interactions between workload assignments that are invisible
    to single-site analysis.

\end{description}

The FSOR is the \emph{intersection} of the feasible sets
defined by each constraint class. This intersection structure
has two immediate consequences. First, relaxing any single
constraint expands the FSOR monotonically in the set-inclusion
sense: a less stringent carbon threshold admits more sites, a
relaxed latency objective admits more placements, a higher link capacity admits more routing configurations. This expansion may itself be discontinuous --- re-admitting a previously excluded
site can restore feasibility for a large number of workload
configurations simultaneously. Second, tightening any single
constraint contracts the FSOR, and the contraction may be
disproportionate: a marginal tightening of the carbon threshold
that eliminates one high-carbon site may remove a large fraction
of the FSOR if that site was the only feasible host for several
latency-constrained workloads. The asymmetry between these two
consequences---smooth relaxation benefits versus potentially
catastrophic tightening effects---has direct implications for
infrastructure investment planning and regulatory threshold
design, as discussed in~\S\ref{sec:sovereignty}.

\subsubsection{Dynamic Evolution}
\label{subsec:fsor_dynamics}

The FSOR is not a static property of infrastructure but a
time-varying set whose boundary evolves continuously, driven by
the dynamics of $\boldsymbol{\theta}(t)$. Three distinct
timescales govern this evolution, each associated with a
different subset of the infrastructure parameters and a
different planning response.

\begin{description}

  \item[Sub-hourly: grid carbon dynamics.]
    Grid carbon intensity $\gamma_i(t)$ varies with generation
    dispatch and demand fluctuations, sometimes by a factor of
    two or more within a single
    day~\cite{radovanovic2021carbon,cote2025locational}. During
    periods of high renewable output, sites previously excluded
    by the carbon gate re-enter the feasible set; during evening
    demand peaks or low-wind periods, they may be excluded
    again. The FSOR therefore expands and contracts on
    sub-hourly timescales, creating what may be termed
    \emph{temporal arbitrage opportunities}: windows during
    which the FSOR is larger than average, into which
    delay-tolerant batch workloads can be shifted to exploit
    lower-carbon grid conditions without violating their relaxed
    latency budgets. Network link utilization also fluctuates
    on this timescale and can transiently tighten the routing
    feasibility component of the FSOR.

  \item[Daily to seasonal: water availability and cooling
    efficiency.]
    Water permit limits $\bar{\Omega}_i(t)$ and ambient cooling
    efficiency $\omega_i(t)$ evolve on daily and seasonal
    timescales, driven by temperature, humidity, and
    hydrological conditions. Summer months in water-stressed
    regions produce systematic FSOR contraction as permit
    headroom narrows and evaporative cooling draws more water
    per unit of IT load. Unlike carbon intensity fluctuations,
    this contraction is largely predictable from climate and
    hydrometeorological data, making it amenable to anticipatory
    capacity planning rather than reactive scheduling, as
    discussed in~\S\ref{subsec:endowments}.

  \item[Annual to decadal: regulatory and climate trajectories.]
    Policy thresholds $\bar{\Gamma}_i$ evolve as regulatory
    frameworks tighten over time, as contemplated in the
    trajectory of EU digital infrastructure sustainability
    targets~\cite{EuropeanCommission2021DC} and the national AI strategies that couple sustainability obligations to
    infrastructure investment~\cite{EuropeanCommission2021AI}.
    Two structural forces operate in opposing directions on this timescale. Grid decarbonization reduces $\gamma_i(t)$
    systematically over multi-year horizons as fossil generation is displaced by renewables, expanding the carbon-eligibility component of the FSOR. Climate change, by contrast, shrinks the water-availability envelope $\bar{\Omega}_i(t)$ over
    decadal timescales through longer and more severe dry
    seasons~\cite{IPCC2021WG1}, contracting the
    water-feasibility component. The net trajectory of the FSOR depends on the relative pace of these two forces and will vary significantly by region.

\end{description}

This multi-timescale structure implies that the FSOR should be
evaluated not as a single snapshot but as a \emph{trajectory}:
a time-varying feasibility boundary whose evolution reflects
the joint dynamics of the grid, climate, and regulatory
environment. Infrastructure investment decisions---site
selection, cooling technology, renewable procurement, and
network topology---alter this trajectory and should therefore
be evaluated against projected FSOR evolution over the intended
operational lifetime of the facility, not only against current
conditions~\cite{Moret2019}. A region whose FSOR is adequate
today may face binding constraints within a planning horizon as
carbon thresholds tighten and water stress intensifies;
conversely, grid decarbonization may expand the FSOR
sufficiently to admit workload classes that are currently
infeasible. The multi-period investment planning implications
of this trajectory are identified as a direction for future
work in~\S\ref{subsec:limitations}.

\subsubsection{Two Sovereignty-Limiting Mechanisms}
\label{subsec:sovereignty_mechanisms}

Two structural mechanisms dominate FSOR contraction and merit
explicit treatment. Both are illustrated in
Figure~\ref{fig:fsor},%
\footnote{Figure~\ref{fig:fsor} is introduced in
\S\ref{subsec:fsor_metric}.} which shows how each mechanism produces a qualitatively distinct pattern of feasibility loss across three representative infrastructure profiles.

The first is \emph{sustainability-driven fragmentation}. As
carbon or water thresholds tighten, sites are removed from the
feasible set discretely. When the remaining compliant sites are
geographically sparse or poorly connected by the optical
network, the feasible placement--routing space fragments:
workload configurations that required co-placement or
low-latency routing across multiple sites may find no feasible
assignment within the reduced site set, even though each
remaining site individually satisfies all constraints.
Fragmentation reduces not only the number of feasible
configurations but also their robustness: the remaining
feasible assignments may be sensitive to small workload
additions or telemetry changes that push the active
configuration outside the FSOR, leaving little margin for
demand growth or constraint tightening.

The second is the \emph{green-but-far effect}, introduced in
\S\ref{subsec:tension} as a constraint interaction and examined
here from the perspective of its sovereignty consequences.
Sustainability-compliant sites may be geographically remote
from demand sources. The propagation delay floor imposed by the
optical network---irreducible below the speed-of-light
bound~\cite{singla2014}---means that latency-sensitive workloads
cannot use these sites regardless of their sustainability
profile. The FSOR boundary for latency-sensitive workload
classes is therefore determined jointly by the sustainability
eligibility of sites and their geographic proximity to demand.
These two criteria may be in systematic tension in regions
where clean-energy resources are concentrated far from
population centers, a geographic pattern that is empirically
well-documented for wind and solar generation in Europe and
North America~\cite{Worman2024}.

\begin{remark}[Asymmetric contraction]
When these two mechanisms interact---sustainability thresholds
fragment the site set, and the remaining compliant sites are
remote---the FSOR may contract to empty for latency-sensitive
workloads while remaining non-empty for delay-tolerant batch
workloads. This asymmetric contraction is not merely a
scheduling inconvenience: it is the formal expression of a
sovereignty constraint. Sovereignty over a class of AI
workloads requires the ability to operate that class locally
under prevailing physical and regulatory conditions. If
latency-sensitive workloads fall outside the FSOR while batch
workloads remain inside it, the infrastructure cannot sustain
real-time AI services under local control---regardless of the
compute capacity nominally available. Identifying which workload
classes fall inside and outside the FSOR under a given
telemetry snapshot is therefore a direct, operationally
grounded measurement of the scope and limits of sovereign AI
capacity.
\end{remark}

\subsubsection{Infeasibility as a Sovereignty Signal}

When the active workload set $\mathcal{W}(t)$ lies outside the FSOR---that is, when
$\Phi\bigl(\mathcal{W}(t),\,\boldsymbol{\theta}(t)\bigr) =
\emptyset$---no valid placement--routing pair exists under
current conditions. The solver returns an
\emph{infeasibility certificate}: a minimal description of the constraints whose simultaneous binding renders the problem unsolvable, formally corresponding to an Irreducible Infeasible Set (IIS) of the constraint
system~\cite{Chinneck2008}. Rather than returning a
constraint-violating placement, the framework surfaces this
certificate directly to the operator. This is a deliberate
design choice: a policy-violating assignment would conceal the sovereignty gap behind a technically invalid solution, whereas the infeasibility certificate makes the gap explicit and actionable.

Infeasibility should be interpreted as a telemetry-grounded
signal with two qualitatively distinct readings, corresponding to the sub-hourly and structural timescales identified in~\S\ref{subsec:fsor_definition}.

The first is \emph{transient infeasibility}: the FSOR is
non-empty at other points in its temporal trajectory, and the current snapshot reflects a temporary period of elevated carbon intensity, water stress, or link congestion. The appropriate operational response depends on workload class: delay-tolerant batch workloads should be deferred until a forecast FSOR expansion creates a compliant scheduling window, while portable workloads with imminent deadlines may be pre-emptively migrated to compliant sites ahead of the constraint binding. Both responses rely on the forecast-based state estimation described in \S\ref{subsec:telemetry}, which provides short-horizon predictions of
$\boldsymbol{\theta}(t + \Delta)$ that allow the orchestrator to anticipate FSOR contraction before it
occurs~\cite{radovanovic2021carbon}.

The second is \emph{structural infeasibility}: the FSOR is
empty or severely contracted across all telemetry snapshots
within the planning horizon, reflecting a fundamental and
persistent mismatch between workload requirements and
infrastructure endowment. This is not a scheduling problem but an infrastructure investment problem. The infeasibility
certificate provides a precise characterization of the gap:
which constraints are simultaneously binding, which sites are excluded and on what grounds, and what minimum infrastructure additions or policy adjustments would restore feasibility.
Concretely, the IIS identifies whether the binding limitation is carbon eligibility (too few low-carbon sites), water headroom (permit limits too tight for the workload power profile), latency admissibility (no compliant site within the propagation-delay radius), or network capacity (insufficient link bandwidth to route the required traffic). Each binding constraint points to a different investment lever.

Used in this way, the FSOR and its infeasibility boundary
become a \emph{planning instrument}: a quantitative,
telemetry-grounded basis for sovereign AI infrastructure
investment that is anchored in physical and regulatory
constraints rather than in abstract capacity targets. The
sovereignty implications of this planning role are developed
in~\S\ref{sec:sovereignty}, where the FSOR is used to
characterize the structural asymmetries in AI infrastructure
capacity across regions with different endowments.
\section{Agentic Control Architecture}
\label{sec:architecture}
\subsection{Overview}
\label{subsec:architecture_overview}

\begin{figure}[t]
  \centering
  \includegraphics[width=\linewidth]{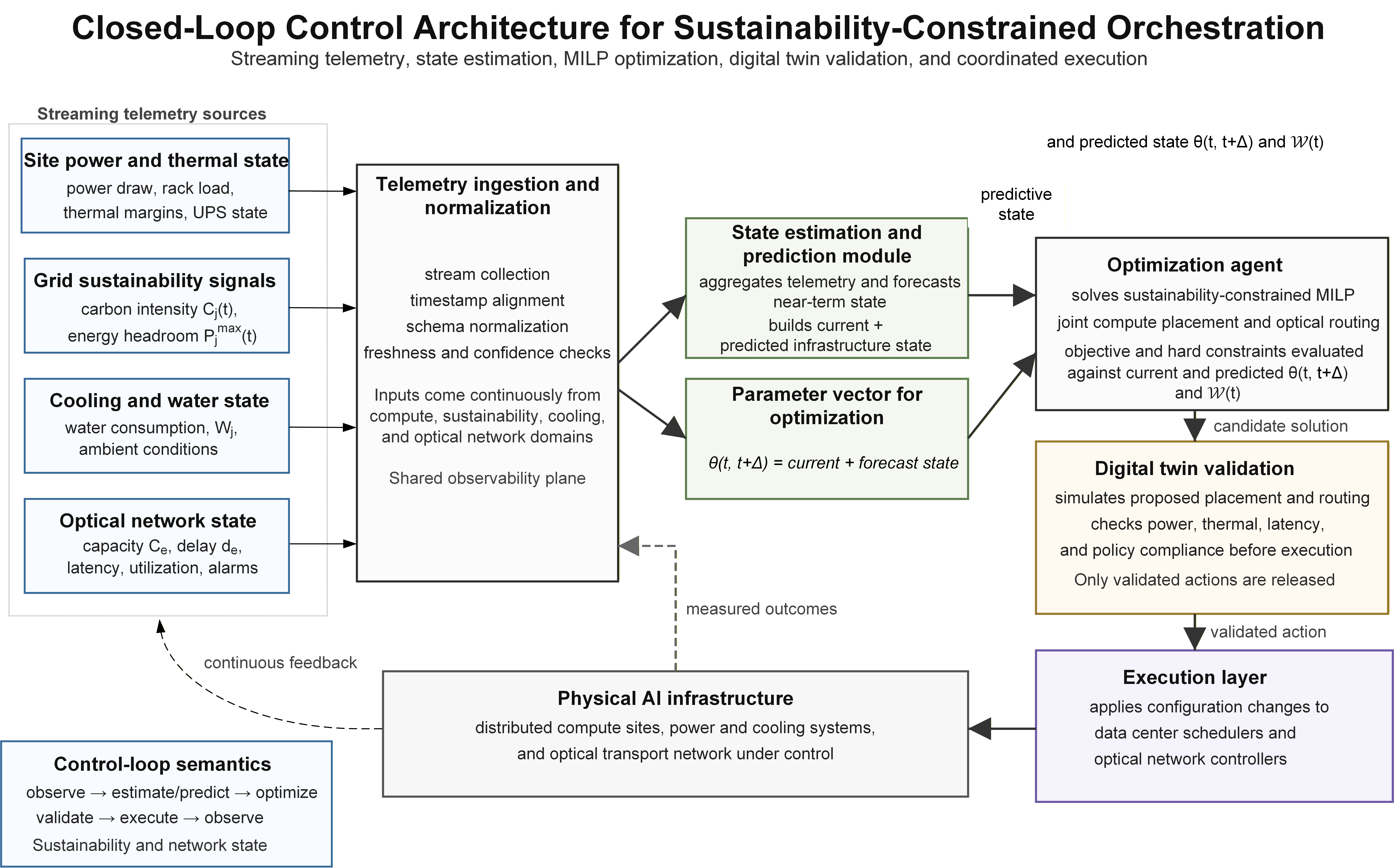}
  \caption{Closed-loop control architecture for sustainability-constrained AI infrastructure orchestration. Streaming telemetry from compute, energy, cooling, and optical network domains is ingested and normalized to provide a unified observability plane. A state estimation and prediction module derives both the current and a short-horizon forecast
    representation of infrastructure state, which jointly parameterize the MILP optimization. The optimization agent computes placement and routing decisions subject to sustainability and latency hard constraints. A
    digital twin validates candidate solutions against power, thermal, latency, and policy compliance before execution. The execution layer applies configuration changes to data center schedulers and optical network controllers, while continuous telemetry feedback closes the loop, forming a predictive closed control system.}
  \label{fig:control-loop}
\end{figure}

The optimization problem formulated in \S\ref{sec:problem} is not solved once in isolation: it is embedded in a closed-loop control system that continuously monitors the physical AI infrastructure and acts on it at each telemetry cycle. The architecture, illustrated in Figure~\ref{fig:control-loop}, realizes sustainability-constrained AI infrastructure operation as a cross-domain control system spanning three physical layers---compute, optical network, and energy/cooling---and five functional stages that execute in sequence within each control cycle: telemetry ingestion and normalization, state estimation and prediction, optimization, digital twin validation, and execution. The loop closes through continuous telemetry feedback from the physical infrastructure, enabling adaptive, constraint-aware orchestration under dynamic operating conditions.

\subsubsection{Control-loop semantics}
The six-phase semantics of the control loop are:
\textsc{observe} $\to$ \textsc{estimate/predict} $\to$ \textsc{optimize}
$\to$ \textsc{validate} $\to$ \textsc{execute} $\to$ \textsc{observe}.
Each phase has a well-defined input--output contract with the adjacent phases, and the loop as a whole implements a receding-horizon control strategy: at each cycle, the optimization is solved over the current and predicted state, a validated action is executed, and the system re-observes before solving again. This structure is common in model predictive control
(MPC)~\cite{Rawlings2009MPC} and has been adapted here to the joint compute--network--energy problem, where the ``plant'' is the distributed AI infrastructure and the ``model'' is the MILP formulation of \S\ref{subsec:opt_problem} evaluated
against the parameter vector $\boldsymbol{\theta}(t, t+\Delta)$.

\subsubsection{State representation}
The system state is represented by the parameter vector
$\boldsymbol{\theta}(t, t+\Delta)$, which encodes both the current
infrastructure conditions at time $t$ and a short-horizon forecast over
the interval $[t, t+\Delta]$ derived from telemetry streams and predictive
models. Concretely, $\boldsymbol{\theta}(t, t+\Delta)$ extends the
per-site tuple $\{P_i(t), \gamma_i(t), \omega_i(t), \bar{\Gamma}_i,
\bar{\Omega}_i(t)\}_{i \in \mathcal{S}}$ defined in
\S\ref{subsec:infra_model} with forecast trajectories for each parameter over the prediction horizon $\Delta$. The active workload set $\mathcal{W}(t)$ represents the workloads present in the system at time $t$ that are subject to placement and routing decisions in the current cycle; workloads that arrive or depart between cycles are handled through incremental re-optimization rather than full problem reconstruction.

The dual role of telemetry is a defining architectural feature. Telemetry outputs are used simultaneously to construct the estimated and predicted
state $\boldsymbol{\theta}(t, t+\Delta)$ and to parameterize the
optimization problem directly---as the values of $\gamma_i(t)$,
$\omega_i(t)$, $P_i(t)$, $C_{ij}$, and $d_{ij}$ that appear in
constraints~\eqref{eq:carbon_gate}--\eqref{eq:flow_conservation}. This dual use reflects the role of telemetry as both an observability layer and a decision input: the same data stream that informs the operator about current infrastructure conditions also determines the boundary of the FSOR at each cycle.

\subsubsection{Cross-domain observability}
A prerequisite for the closed-loop architecture is a shared observability plane that integrates telemetry from four physically distinct domains, each with its own instrumentation standards, update rates, and schema
conventions. As shown in Figure~\ref{fig:control-loop}, these are: site power and thermal state (power draw, rack load, thermal margins, UPS state); grid sustainability signals (carbon intensity $\gamma_j(t)$, energy headroom $P_j^{\max}(t)$); cooling and water state (water consumption $W_j$, ambient conditions); and optical network state (capacity $C_e$, delay $d_e$, utilization, alarms). The telemetry ingestion and normalization stage is responsible for stream collection, timestamp alignment across domains, schema normalization, and freshness and confidence checks that flag stale or low-confidence readings before they propagate into the state estimator and optimization. Without this normalization layer, heterogeneous update rates across domains would introduce temporal inconsistencies in the parameter vector $\boldsymbol{\theta}(t, t+\Delta)$---for example, a carbon intensity reading from a slow-updating grid API paired with a freshly observed power headroom---that could cause the optimization to evaluate constraints against an internally inconsistent system state.

\subsubsection{Predictive versus reactive control}
A purely reactive architecture---one that responds only to the current telemetry snapshot $\boldsymbol{\theta}(t)$---is insufficient for sustainability-constrained orchestration for two reasons. First, the timescales of constraint activation differ across domains: carbon intensity can change within minutes, water permit limits activate over hours to days, and link congestion evolves on timescales comparable to the optimization and execution cycle. A controller that observes only the current state will systematically lag behind fast-moving constraints, executing placements that were valid at observation time but invalid at execution time. Second, several sustainability-improving actions---pre-emptive workload migration to sites whose carbon intensity is forecast to fall, reservation of routing capacity ahead of predicted congestion---are only available to a controller that can plan ahead. The forecast horizon $\Delta$ in
$\boldsymbol{\theta}(t, t+\Delta)$ is the architectural mechanism that enables this anticipatory behavior: the optimization evaluates hard constraints not only against the current state but against the predicted state at execution time, reducing the probability of constraint violations that arise from the latency between observation and action.

\subsubsection{Digital twin as a safety layer}
The digital twin validation stage occupies the position between the optimization agent and the execution layer and serves as a mandatory safety check. Every candidate solution produced by the optimization agent is simulated in the digital twin before any configuration change is applied to the physical infrastructure. The twin checks power balance, thermal margins, end-to-end latency, and regulatory policy compliance under the
proposed placement and routing. Only validated actions are released to the execution layer; solutions that fail validation are returned to the optimization agent with a constraint tightening that excludes the offending configuration, triggering re-optimization. This reject-and-retry loop
ensures that the physical infrastructure is never exposed to an
unvalidated action, at the cost of an additional round-trip latency within the control cycle. The design choice to place validation between optimization and execution---rather than relying on the optimization model alone to guarantee constraint satisfaction---reflects the inherent gap between the MILP's abstracted model of the infrastructure and the full complexity of the physical system, including thermal dynamics, transient power spikes, and optical signal-quality margins that are not captured in the placement and routing formulation.

\subsubsection{Execution and feedback closure}
The execution layer translates validated placement and routing decisions into concrete configuration changes applied to data center schedulers and optical network controllers. The feedback path closes the loop: once the physical infrastructure has responded to the configuration change, its
updated state is immediately observable through the telemetry streams, and the next control cycle begins. The feedback arrow in
Figure~\ref{fig:control-loop} carries \emph{measured outcomes}---the actual power draw, carbon consumption, water draw, and link utilization observed after execution---back to the telemetry ingestion stage, where they serve two purposes: updating the state estimate for the next cycle and providing a ground-truth signal for evaluating forecast accuracy and detecting model drift in the state estimation and prediction module. This last use is essential for maintaining the predictive fidelity of $\boldsymbol{\theta}(t, t+\Delta)$ over time: a state estimator whose forecasts systematically diverge from measured outcomes will generate parameter vectors that cause the optimization to evaluate constraints against an increasingly inaccurate model of the physical system.

\subsection{Telemetry Collection}
\label{subsec:telemetry}

Telemetry collection is the observability foundation of the closed-loop architecture~\cite{cruzes2026telemetry}. Every constraint in the optimization
problem~\eqref{eq:carbon_gate}--\eqref{eq:flow_conservation} is
parameterized by quantities that must be measured in real time; the accuracy, freshness, and consistency of those measurements directly determine the fidelity of the FSOR boundary computed at each control cycle. The telemetry layer is therefore not a peripheral data-collection mechanism but a first-order determinant of control quality.

Streams are collected from four physically distinct domains, each with its own instrumentation ecosystem, update rate, and schema conventions. A shared ingestion pipeline performs timestamp alignment, schema normalization, and data quality enforcement before any stream is admitted to the state estimator. The remainder of this subsection describes each domain in turn, then addresses the cross-cutting concerns of schema normalization, freshness enforcement, and graceful degradation under partial observability.

\subsubsection{Compute and Power Telemetry}

Site power and thermal state is collected at rack and accelerator
granularity via standard data center management interfaces, including \textsc{ipmi}, Redfish, and vendor-specific baseboard management controller (\textsc{bmc}) APIs~\cite{DMTF2024Redfish}. The primary variables are:

\begin{itemize}
  \item \textbf{Per-rack power draw} \textnormal{[kW]}: the instantaneous
    electrical load on each rack, used to compute the site-level power headroom $P_i(t) - \sum_{k} p_k x_{ik}$ against the
    constraint~\eqref{eq:power};
  \item \textbf{Accelerator utilization} \textnormal{[\%]}: GPU/TPU
    compute utilization per device, used to infer available capacity for workload migration and to detect underutilized resources that could absorb additional placement;
  \item \textbf{Thermal sensor readings} \textnormal{[°C]}: intake air
    temperature, chip junction temperature, and coolant supply/return delta, used to assess thermal headroom and flag sites approaching thermal throttling thresholds;
  \item \textbf{UPS state}: battery charge level, bypass status, and estimated backup runtime, used to assess site availability under grid instability conditions.
\end{itemize}

Power and thermal telemetry are among the fastest-updating streams in the architecture, with sub-minute update rates typical for rack-level power meters. The digital twin validation stage relies on these readings to verify that a proposed placement does not drive any site into thermal saturation or exceed its contracted power envelope, making high-freshness power telemetry a prerequisite for safe execution.

\subsubsection{Grid Sustainability Telemetry}

Sustainability telemetry covers three sub-streams: grid carbon intensity, on-site renewable generation and storage state, and cooling water consumption.

\subsubsection{Grid Carbon Intensity}

Marginal grid carbon intensity $\gamma_i(t)$
\textnormal{[gCO$_2$eq/kWh]} is sourced from commercial grid
signal providers such as Electricity Maps or
WattTime~\cite{cote2025locational}, which aggregate generation
dispatch data, fuel mix reports, and flow-tracing models to
produce real-time and short-horizon forecast estimates at the
bidding-zone or balancing-area level. The choice between
marginal and average emissions factors has a material effect on
the signal used to parameterize the carbon
constraint~\eqref{eq:carbon_gate}~\cite{ElectricityMaps2023}:
marginal factors reflect the actual incremental emitter
dispatched to serve additional load and are theoretically correct
for placement decisions that alter load at the margin, while
average factors reflect the overall generation mix and are more
stable but less sensitive to short-term dispatch
changes~\cite{ElectricityMaps2023}. The framework admits either
convention as a configuration parameter; the choice should be
made explicit and held constant across sites to avoid comparing
incommensurable signals.

\subsubsection{On-Site Generation and Storage}

Where sites are equipped with behind-the-meter renewable
generation or battery storage, their effective carbon intensity
and power headroom are functions of on-site generation state in
addition to the grid signal. The net power headroom available to
the placement constraint~\eqref{eq:power} is
$P_i(t) = P_i^{\mathrm{grid}}(t) + P_i^{\mathrm{gen}}(t) +
P_i^{\mathrm{batt}}(t)$, where the latter two terms are sourced
from the energy management system of the on-site generation and
storage assets. Sites with significant on-site renewable capacity
may remain inside the carbon-eligibility set defined
by~\eqref{eq:carbon_gate} even when the grid signal $\gamma_i(t)$
exceeds $\bar{\Gamma}_i$, if their effective blended carbon
intensity falls below the threshold.

\subsubsection{Cooling Water Consumption}

Water consumption $W_j(t)$~\textnormal{[L/h]} and make-up water
flow are monitored via building management system
(\textsc{bms}) interfaces. Ambient environmental
conditions---intake air temperature, relative humidity, wet-bulb
temperature---are collected from on-site weather stations and
used to model cooling system efficiency and forecast evaporative
water draw under predicted ambient conditions over the horizon
$\Delta$. These inputs feed directly into the water constraint
evaluation~\eqref{eq:water_capacity} and into the seasonal FSOR
contraction dynamics discussed later in~\S\ref{subsec:endowments}.
\subsubsection{Optical Network Telemetry}

Optical transport network state is collected via OpenConfig YANG models or vendor streaming telemetry interfaces (\textsc{gnmi}/\textsc{grpc}) \cite{Shakir2018gNMI}, which expose per-channel and per-link performance monitoring data at update rates of seconds to minutes. The primary variables are:

\begin{itemize}
  \item \textbf{Channel capacity} $C_e$ \textnormal{[Gbps]}: the
    provisioned and available transmission capacity on each directed edge,
    used to parameterize the link capacity
    constraint~\eqref{eq:link_capacity};
  \item \textbf{Propagation delay} $d_e$ \textnormal{[ms]}: the one-way
    fiber propagation delay on each link, used to evaluate the latency
    constraint~\eqref{eq:latency_slo} via~\eqref{eq:latency}. This
    quantity is physically stable---it is determined by fiber route geometry
    and refractive index---and requires re-measurement only when route
    changes or protection switching events alter the active fiber path;
  \item \textbf{Link utilization} \textnormal{[\%]}: the fraction of
    capacity currently occupied by active flows, used to compute available
    headroom for new traffic demands and to detect congestion conditions
    that may invalidate routing assumptions;
  \item \textbf{Signal quality indicators}: optical signal-to-noise ratio
    (\textsc{osnr}), pre-forward-error-correction bit error rate
    (\textsc{pre-fec ber}), and chromatic dispersion margin, used by the
    digital twin to assess whether a proposed routing is physically
    realizable on the optical layer without degrading signal integrity;
  \item \textbf{Alarm state}: active faults, degraded spans, and
    protection switching events, used to remove affected links from the
    routing graph before optimization.
\end{itemize}

Optical network telemetry introduces a domain-specific challenge: the
logical routing graph $\mathcal{G} = (\mathcal{S}, \mathcal{E})$ used in the optimization is an abstraction of a multi-layer optical network whose actual capacity and delay properties depend on wavelength assignment, amplifier gain settings, and protection topology. Maintaining consistency between the optimization's graph abstraction and the physical optical layer
requires that the telemetry pipeline translates physical-layer performance monitoring data into the logical-layer parameters $C_e$ and $d_e$ used in the MILP. Significant discrepancies between the two---for example, a link whose effective capacity is reduced by signal degradation but whose provisioned capacity is unchanged in the logical model---are a source of digital twin validation failures and should trigger a re-parameterization of the graph.

\subsubsection{Workload Telemetry}

Workload telemetry is collected from the data center orchestration layer (Kubernetes control plane, \textsc{slurm} job scheduler, or equivalent) and from application-level monitoring. The primary variables
are~\cite{cruzes2026telemetry}:

\begin{itemize}
  \item \textbf{Active workload identifiers and current placement}: the set
    $\mathcal{W}(t)$ and the current assignment $\mathbf{x}(t)$, used to
    initialize the optimization with the current state and to identify workloads that may be candidates for migration;
  \item \textbf{Traffic demand} $\rho_k(t)$ \textnormal{[Gbps]}: the
    observed inter-site traffic generated by each workload, used to
    parameterize the flow conservation constraint~\eqref{eq:flow_conservation}
    and the link capacity constraint~\eqref{eq:link_capacity};
  \item \textbf{Power demand} $p_k(t)$ \textnormal{[kW]}: the measured
    power draw of each workload, which may differ from the nominal $p_k$
    used in placement planning due to load variation;
  \item \textbf{Service-level compliance state}: per-workload latency
    measurements against the \textsc{slo} $\lambda_k$, used to detect
    active violations that may trigger re-optimization outside the regular
    cycle interval;
  \item \textbf{Portability and migration state}: whether a workload is
    currently migrating ($\mu_k$ in transition), the size of its
    transferable state $\sigma_k$, and the estimated completion time of any
    in-flight migration, used to avoid issuing conflicting placement
    decisions for workloads that are mid-migration.
\end{itemize}

\subsubsection{Ingestion Pipeline: Normalization, Freshness, and
  Graceful Degradation}

All four telemetry streams are timestamped at the source and admitted through a shared ingestion pipeline before any reading is made available to the state estimator. The pipeline enforces three cross-cutting properties.

In Schema normalization,  each domain exposes data in domain-specific formats and units. The ingestion pipeline applies a schema normalization layer that maps all streams to a common internal representation with standardized units, parameter names, and site identifiers. This normalization is a prerequisite for the state estimator's ability to join readings across domains---for example, combining the carbon intensity signal from Electricity Maps with the power headroom reading from the \textsc{bmc} and the link capacity from OpenConfig into a single per-site parameter vector
$\boldsymbol{\theta}_i(t)$.

In Timestamp alignment, streams from different domains have heterogeneous update rates: power meters may update every 30 seconds, carbon intensity signals every 5 minutes, and water consumption readings every hour. The ingestion pipeline
aligns all streams to a common control cycle timestamp using
last-known-good interpolation for slowly varying quantities and
forward-hold for quantities with hard physical meaning (alarm states, UPS bypass). Misaligned timestamps that exceed a configurable tolerance are flagged as potentially inconsistent and trigger a confidence reduction on the affected parameter before it enters the state estimator.

In Freshness enforcement and graceful degradation, 
telemetry older than a configured maximum age $\tau_{\max}$ for each parameter is treated as stale and triggers one of three responses, depending on the criticality of the affected parameter:
\begin{enumerate}
  \item \textbf{Substitution from forecast}: for slowly varying parameters
    such as water consumption and ambient temperature, the state estimator
    substitutes the most recent forecast value with a widened uncertainty
    interval, allowing optimization to proceed with reduced confidence;
  \item \textbf{Conservative bound}: for safety-critical parameters such as
    power headroom and link capacity, the pipeline substitutes a
    conservative lower bound---for example, the minimum observed value over
    the preceding hour---preventing the optimization from assuming
    availability that cannot be verified;
  \item \textbf{Optimization hold}: for parameters with no safe substitute
    ---such as alarm state on a link that may be in active protection
    switching---the pipeline signals the orchestrator to hold optimization
    until fresh telemetry is received, preventing decisions on an
    irrecoverably stale system state.
\end{enumerate}
This three-tier degradation policy ensures that the control loop continues to operate safely under partial observability, while preventing unsafe actions that could arise from acting on stale or missing telemetry. It also provides a principled interface between the telemetry layer and the state estimation module: rather than passing raw readings with unknown freshness,
the ingestion pipeline delivers a parameter vector annotated with
per-parameter confidence levels that the state estimator can use to weight its fusion of telemetry and forecast inputs.

\subsection{State estimation}

A state estimator aggregates incoming telemetry into a consistent
infrastructure snapshot. Slowly varying parameters such as water intensity and physical link delays are smoothed over time to suppress measurement noise. Rapidly varying parameters such as grid carbon intensity and available power capacity use the most recent telemetry reading directly. Each parameter estimate carries a confidence flag reflecting data age and source reliability. When confidence is low, the optimizer may apply conservative bounds---using a pessimistic carbon estimate rather than the point estimate---trading solution quality for robustness against measurement uncertainty.

Recent advances in multivariate time-series forecasting for cloud
infrastructure suggest that state estimation should incorporate predictive components, capturing both current conditions and near-term resource dynamics. Hybrid deep learning models have demonstrated the ability to model interdependencies between compute, memory, and network resources, enabling more accurate and proactive orchestration decisions~\cite{sallam2025temposight}.
In this context, the system state is not only an instantaneous snapshot, but a short-horizon forecast that informs feasibility-aware optimization under dynamic conditions.

\subsection{Optimization agent}

Given the current and predicted infrastructure state and active workload set, the optimization agent first eliminates infeasible site--workload pairs by applying carbon and water constraints.  This pre-filtering step reduces problem size before the solver runs and makes the hard-constraint semantics explicit in the
problem structure.  The agent then solves the joint placement and routing problem, returning an optimal assignment if the problem is feasible or raising an infeasibility flag if no valid placement exists under current conditions.

The agent operates on a periodic cycle aligned with the update frequency of carbon intensity signals, typically in the range of five to fifteen minutes. This cadence balances responsiveness to changing grid conditions against the overhead of repeated optimization and execution.

\subsection{Digital twin validation}

Before any solution is applied to the physical infrastructure, it
is validated against a digital twin, a continuously synchronized
model of the system~\cite{cruzes2026sovereignty}. The digital twin
evaluates the proposed placement against four categories of
constraints that are difficult to encode precisely in the
optimization model: thermal margins (does the new workload
allocation remain within the facility's cooling capacity under
current ambient conditions?); power stability (do the dynamic
fluctuations implied by the new allocation remain within UPS and
transformer headroom?); network congestion (does the proposed
routing create hotspots or violate restoration margins on shared
optical links?); and policy compliance (does the proposed solution satisfy data locality requirements and cross-border traffic restrictions?).

If any check fails, the digital twin returns a constraint violation
report; the optimization agent adds a corresponding cut and
re-solves. This correction loop repeats until a valid solution is
found or the cycle time budget is exhausted. In the latter case,
the current placement is retained and an operator alert is raised.
This validation step ensures that optimization outputs are not only
mathematically feasible, but operationally safe under real-world
conditions that are not fully captured in the optimization model.

\subsection{Execution}

Validated solutions are applied atomically across two control planes.  On the \emph{compute plane}, workload placement instructions are issued to the cluster scheduler, specifying target node affinities, power caps, and priority levels.
On the \emph{network plane}, routing instructions are issued to the optical network controller via the path computation element interface, updating label-switched path assignments and optical channel configurations. The two planes are coordinated to apply changes simultaneously, preventing transient states in which workload and routing are inconsistent.

\section{Scenario-Based Evaluation}
\label{sec:evaluation}

\subsection{Evaluation approach}

The framework is evaluated through scenario-based analysis across
three representative infrastructure configurations. The scenarios
are designed to isolate distinct sustainability-limiting mechanisms:
carbon intensity variation, spatial carbon--latency tradeoffs, and
seasonal water stress. Results are comparative and qualitative in
nature, demonstrating the structural advantages of joint optimization
relative to decoupled and unconstrained baselines. They are not
intended as universal performance claims---actual reductions in real
deployments will depend on infrastructure characteristics, workload
mix, and prevailing environmental conditions. Accordingly, all
comparisons should be interpreted as directional evidence of relative
behavior rather than as quantitative performance gains.

Three configurations are compared across all scenarios:
\begin{itemize}[leftmargin=*]
  \item \textbf{Baseline:} workloads placed to minimize latency only,
        with no sustainability constraints.
  \item \textbf{Compute-only:} sustainability objective applied to
        placement, but network routing treats all paths as equivalent.
  \item \textbf{Joint (proposed):} full sustainability-constrained
        joint compute--network optimization.
\end{itemize}

\subsection{Scenarios}

\textbf{Scenario A --- Homogeneous continental region}
Five sites distributed across a single continental market with
moderate and relatively uniform grid carbon intensity. Optical
latencies between sites are low. This scenario isolates the benefit
of sustainability-aware placement in a nearly uniform infrastructure,
where environmental gains are achievable without latency tradeoffs.
The joint and compute-only configurations perform similarly,
confirming that network coupling is a minor factor when sites are
well-connected.

\textbf{Scenario B --- Multi-region mixed-carbon grid}
Eight sites spanning two grid zones with strongly differentiated
carbon intensity. Two sites have access to dedicated renewable
generation; others rely on carbon-intensive grid mixes. Optical
latencies vary substantially across zones. This scenario activates
the green-but-far effect (Figure~\ref{fig:green_but_far}): the most
carbon-efficient sites are also the most geographically remote. The
joint optimizer finds placements that balance carbon compliance with
latency feasibility, while the compute-only optimizer places workloads
at low-carbon sites without accounting for the resulting high-latency
routing, leading to service-level violations that require post-hoc
correction.

\textbf{Scenario C --- Water-stressed deployment}
Six sites, a subset of which are located in water-stressed regions
and exceed their water permit thresholds during summer months. Carbon
intensity is moderate. This scenario directly activates the water
constraint and tests FSOR contraction under seasonal stress. During
peak stress periods, the optimization agent declares infeasibility
for a subset of cycles. These infeasibility events correspond to
conditions where water constraints eliminate multiple sites, carbon
constraints eliminate further sites, and the remaining sites lack
sufficient power capacity for the full workload set.

\subsection{Results and interpretation}

Across all scenarios, joint optimization yields lower aggregate
environmental impact relative to both baselines and compute-only
configurations. The difference relative to the compute-only
configuration is most evident in Scenario~B, where decoupling
placement from routing leads the compute-only approach to select
placements that are carbon-efficient but routing-infeasible under
latency constraints. The joint optimizer resolves this tension by
finding placements that are simultaneously carbon-compliant and
latency-feasible, a solution space the compute-only approach cannot
explore.

Latency impact across all configurations is modest. Latency-sensitive
inference workloads are never placed in violation of their
service-level objectives. Any latency increase is absorbed by
delay-tolerant batch analytics workloads, for which temporal
flexibility is an explicit design property. This is consistent with
the latency constraint operating as a hard limit, rather than a soft
trade-off.

In Scenario~C, the infeasibility events observed during peak water
stress highlight an important qualitative behavior. The optimizer
does not return a policy-violating solution; it correctly identifies
that no valid placement exists under current conditions. This behavior
provides operators with precise, actionable information: the
infrastructure cannot sustain the current workload set at the required
constraint levels, and sovereign AI operation requires either expanded
infrastructure capacity or workload shedding. This is categorically
different from a penalty-based formulation, which would return a
``best-effort'' result that silently violates regulatory limits. In
this sense, infeasibility is not an error condition but an explicit
indication that the system is operating outside its feasible sovereign
operating region. The formal structure of these infeasibility
events---and their interpretation as Irreducible Infeasible
Sets---is analyzed in~\S\ref{subsec:hard_vs_penalty_disc}.

The observed behavior is expected to be heterogeneous across regions
and deployment environments. Prior work has shown that the benefits
of AI in energy systems depend on infrastructure maturity, regulatory
conditions, and regional development
levels~\cite{ye2026energyjustice}: the gap between the joint and
compute-only configurations visible in Scenario~B will be larger in
regions where low-carbon sites are geographically concentrated, and
smaller where the grid is relatively uniform as in Scenario~A. The
effectiveness of sustainability-constrained optimization is further
conditioned on the quality of resource forecasting: prediction errors
directly affect scheduling efficiency and energy
utilization~\cite{sallam2025temposight}, implying that the
infeasibility signals observed in Scenario~C are only actionable if
the state estimator can anticipate constraint binding before it
occurs.
\section{Sovereignty Implications}
\label{sec:sovereignty}

\subsection{The FSOR as a sovereignty metric}
\label{subsec:fsor_metric}

\begin{figure*}[t]
  \centering
  \includegraphics[width=\textwidth]{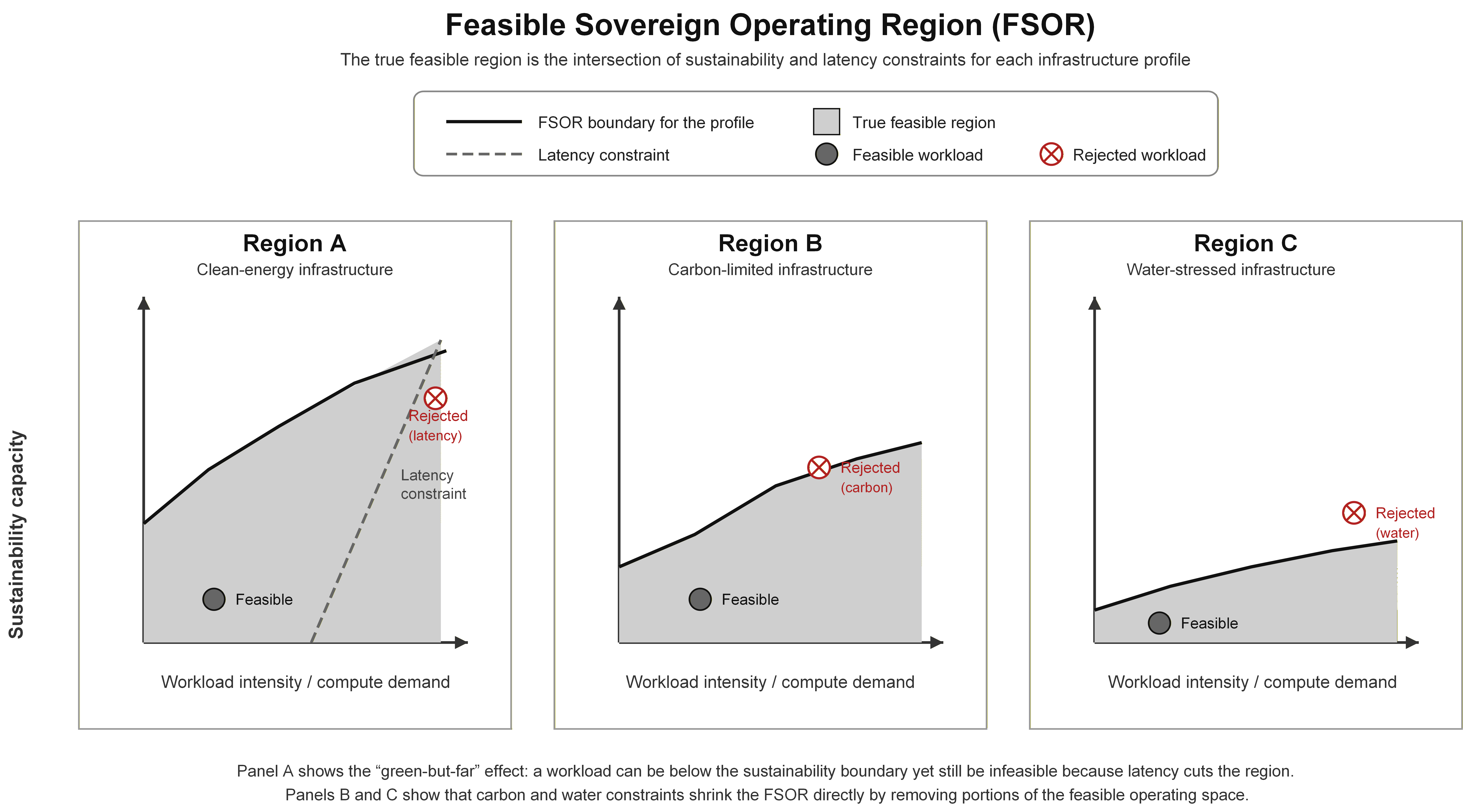}
  \caption{Feasible Sovereign Operating Region (FSOR) under three
    representative infrastructure profiles. In each panel, the shaded area denotes the feasible operating region: the set of workloads that simultaneously satisfy all active constraints, including energy availability, carbon intensity, water usage, and network latency. Region~A (clean energy access) exhibits a larger sustainability envelope, but the latency constraint excludes some otherwise-compliant placements, illustrating the ``green-but-far'' effect. Region~B (carbon-limited grid) shows how stricter carbon thresholds directly contract the feasible
    operating space. Region~C (water-stressed) illustrates further
    contraction driven by cooling and water permit constraints.  The FSOR is the \emph{intersection} of constraint-feasible regions, not a single universal boundary.}
  \label{fig:fsor}
\end{figure*}

Figure~\ref{fig:fsor} illustrates the FSOR concept across three representative infrastructure profiles. Each panel depicts a distinct configuration, with the shaded region representing the feasible operating envelope---the set of workloads that simultaneously satisfy all active constraints under that profile's infrastructure endowment.

The FSOR provides a concrete, measurable definition of sovereign AI capacity. A region's ability to operate AI infrastructure autonomously is bounded by its FSOR: no workload that falls outside the shaded envelope in Figure~\ref{fig:fsor} can be executed under the region's prevailing physical and regulatory constraints. Unlike conventional sovereignty metrics focused on data localization or model ownership rules, the FSOR is expressed entirely in terms of physically observable parameters---power, carbon intensity, water availability, and optical round-trip latency---and is computable from real-time telemetry. It is therefore an \emph{operational} metric rather than a merely conceptual one: it can be monitored, updated as constraints change, and used directly as an input to placement decisions.

While the FSOR defines the physical feasibility of AI operation, the effective realization of this capacity depends on how benefits and constraints are distributed across regions and actors.

Identifying the FSOR establishes what is physically and regulatorily possible; it does not, by itself, determine what is operationally achievable. Effective sovereignty depends equally on the degree of control over the underlying infrastructure. A region may possess a well-defined FSOR and yet be unable to execute within it if the compute, network, or platform resources that realize that envelope are externally owned or governed. Reliance on hyperscale cloud providers, for example, introduces structural dependencies: the provider's scheduling policies, capacity allocation priorities, and jurisdictional obligations---including compelled disclosure under instruments such as the \textsc{cloud} Act---may override the operator's own sustainability and sovereignty requirements~\cite{Daskal2018,Chander2022Sovereignty}. In such cases, the \emph{reachable} subset of the FSOR is strictly smaller than the FSOR itself, and the gap between the two is a direct measure of the infrastructural dependency the region bears. Reducing that gap is, in operational terms, the objective that motivates sovereign AI investment.

Feasibility does not imply equitable access. The physical boundaries represented by the FSOR are not uniformly distributed: regions with abundant renewable capacity, access to submarine cable landing points, and relaxed water-permit regimes face systematically larger feasible envelopes than regions constrained on all three dimensions simultaneously. This structural asymmetry has governance implications beyond the purely technical. Recent work on energy justice shows that AI-enabled systems can simultaneously improve aggregate efficiency while exacerbating distributional inequalities across regions and populations~\cite{ye2026energyjustice}. The FSOR should therefore be read not only as a technical boundary but also as a \emph{structural} boundary that determines which actors can effectively participate in AI infrastructure---and, by extension, which cannot. Policies aimed at broadening sovereign AI capacity must address not only the expansion of individual regional FSORs but also the systematic factors that produce asymmetry across them.

A key feature of the FSOR visible in Figure~\ref{fig:fsor} is that sustainability constraints do not degrade performance gradually as they tighten: they remove entire regions of the feasible space discontinuously. A carbon limit that eliminates a candidate site removes every workload that could have been placed there; a water constraint that activates during a summer stress period removes those sites from the feasible set for the full duration of that period. The FSOR is therefore a set defined by the \emph{intersection} of hard boundaries, not a smooth Pareto frontier along which operators can trade off one objective against another. This discrete structure has a direct practical implication: marginal investments in sustainability compliance---slightly cleaner grid mix, slightly reduced water draw---may yield no benefit at all until a threshold is crossed, at which point the feasible set expands discontinuously. Investment planning for sovereign AI infrastructure should account for this threshold geometry rather than assuming smooth returns to sustainability improvements.

\subsection{Infrastructure Endowments Determine FSOR Boundaries}
\label{subsec:endowments}

The three panels in Figure~\ref{fig:fsor} are not merely illustrative: they represent qualitatively distinct constraint regimes whose interaction shapes the feasible sovereign operating region in ways that cannot be reduced to a single infrastructure bottleneck. Three mechanisms dominate FSOR boundaries across the profiles: access to low-carbon energy, the structural effect of carbon thresholds, and seasonal water availability. Each operates on a different timescale, responds to different policy levers, and produces a different geometry of feasibility loss. Understanding all three is necessary for infrastructure investment planning that aims to expand or defend sovereign AI capacity.

\subsubsection{Clean Energy Access and the Green-But-Far Effect (Region~A)}

Sites with access to low-carbon generation---whether through a predominantly renewable national grid, a power purchase agreement with a dedicated renewable source, or on-site generation---remain inside the feasible set under strict carbon thresholds. Carbon-intensive sites are excluded entirely, regardless of their latency or compute profile. This creates the first boundary mechanism: grid carbon intensity is a hard gate on site eligibility, and it cannot be crossed through compute investment alone. A region whose grid is predominantly fossil-fuelled faces a carbon-limited FSOR that can only be expanded by grid decarbonization, dedicated renewable procurement, or the construction of behind-the-meter generation capacity~\cite{shehabi2024datacenter}. Increasing on-site compute density, improving hardware efficiency, or optimizing scheduling all leave this boundary unchanged.

Region~A also introduces a subtler constraint: the \emph{green-but-far} effect, whose constraint geometry is illustrated in Figure~\ref{fig:green_but_far} (Panel~A) and whose workload-class consequences are mapped in Panel~B of the same figure. A site that is sustainability-compliant in every dimension---low carbon, low water draw, sufficient power headroom---may nonetheless be excluded from the feasible set for latency-sensitive workloads if it is geographically remote from the demand source. As established in \S\ref{subsec:tension}, the propagation delay floor imposed by fiber geometry is irreducible~\cite{singla2014}, meaning that latency-sensitive workloads are spatially bounded around the demand source independently of site sustainability profiles. For inference workloads with tight response-time service-level objectives, this means that the subset of the FSOR accessible to those workloads is spatially bounded around demand center, independently of the sustainability profile of more distant sites. The practical implication is that expanding the FSOR for latency-sensitive workloads requires either demand disaggregation, distributing inference endpoints closer to users, or investment in edge infrastructure that meets sustainability thresholds locally.
Relying on remote clean-energy sites alone will not resolve the tension; the two constraints must be addressed jointly.
Sovereignty policy that focuses exclusively on energy transition without accounting for the latency geometry of AI workloads therefore risks producing infrastructure that is physically compliant but operationally inaccessible for the workloads it is intended to serve.

\subsubsection{Carbon Threshold Geometry and Feasible Set Fragmentation (Region~B)}

Region~B illustrates the structural effect of tightening carbon thresholds on a heterogeneous infrastructure portfolio. As the carbon intensity limit decreases, sites that previously satisfied the constraint cross the threshold and are removed from the feasible set, not partially degraded, but eliminated entirely. This is the threshold geometry discussed in \S\ref{subsec:fsor_metric}: contraction is discontinuous and site-level, not smooth and marginal.

The consequence for operators whose infrastructure spans both low- and high-carbon zones is feasible set \emph{fragmentation}. When a subset of sites is removed by a binding carbon threshold, the remaining feasible placements may form a disconnected or sparsely connected topology. This matters for two reasons. First, workloads with joint placement and routing requirements---for example, a distributed training job that must co-locate data preprocessing and compute, or a federated inference pipeline that must satisfy aggregate latency across multiple endpoints---may find that no feasible assignment exists within the fragmented set, even though individually each remaining site is compliant.
The MILP formulation captures this through the joint satisfaction of placement and routing constraints; the FSOR visualization in Figure~\ref{fig:fsor} projects it into the two-dimensional space of the figure for interpretability.
Second, fragmentation reduces redundancy: fewer compliant sites means reduced ability to absorb failures or demand spikes without violating constraints.

The policy implication is that carbon threshold design is not merely an environmental instrument; it is a de facto infrastructure policy that determines the topology of feasible sovereign compute. Regulators setting carbon intensity limits for public AI infrastructure, as contemplated in emerging EU AI Act guidance and in national AI strategies that couple sustainability obligations to public procurement~\cite{EuropeanCommission2021}, should model the feasibility-set consequences of threshold choices, not only the aggregate emissions reduction they produce. A threshold that is marginally stricter may eliminate a disproportionately large share of feasible sites if those sites cluster near the current limit.

\subsubsection{Seasonal Water Stress and Anticipatory Capacity Planning (Region~C)}

Region~C introduces the third boundary mechanism: water availability. Data centers operating air-cooled or evaporative cooling systems draw substantial volumes of water, and in water-stressed regions this draw is subject to permit limits that tighten during summer months when both cooling demand peaks and river or groundwater levels fall~\cite{Mytton2021}. When a site's water draw exceeds its permitted allocation, it exits the feasible set for the duration of the stress period, contracting the FSOR seasonally in a manner that is predictable from climate data and cooling efficiency parameters.

The temporal structure of water stress distinguishes it from the other two mechanisms. Carbon intensity varies on sub-hourly timescales, driven by generation dispatch and demand fluctuations; latency constraints are effectively static on planning horizons relevant to infrastructure investment.
Water stress, by contrast, evolves on seasonal timescales, weeks to months, with a climatological signal that can be forecast with reasonable accuracy months in advance using hydrometeorological models. This predictability makes water-driven FSOR contraction amenable to \emph{anticipatory} planning: an operator can pre-position workloads away from water-stressed sites before permit thresholds bind, reserve capacity at alternative sites during summer months, or negotiate interruptible cooling agreements that partially substitute air-side economization for evaporative cooling when water is scarce.

The structural trend, however, is adverse. Climate projections for many high-compute regions---including Southern and Central Europe, the southwestern United States, and parts of East Asia---point to longer and more severe dry seasons, increasing the fraction of the year during which water-stress constraints are active~\cite{IPCC2021WG1}. This means that the seasonal FSOR contraction in Region~C is not a static planning parameter but a \emph{shrinking envelope} over multi-year horizons. Sovereign AI capacity planning must therefore treat water availability not as a fixed endowment but as a time-varying constraint whose trajectory is coupled to climate risk.
Infrastructure investment decisions---site selection, cooling technology choice, water recycling systems, co-location with waste-heat recovery---should be evaluated against projected water availability over the intended operational lifetime of the facility, not only current permit conditions.

Finally, it is worth noting that the three mechanisms interact. A site under Region~C conditions that also lies near the carbon threshold of Region~B may be eliminated from the feasible set on two independent grounds simultaneously during summer, while remaining accessible during winter. An operator managing a portfolio that spans all three constraint regimes faces a feasible set that contracts and expands non-uniformly across the calendar year, across the grid's generation mix, and across its routing topology. The FSOR framework, by making these intersecting boundaries explicit and computable from telemetry, provides the basis for the kind of joint, constraint-aware capacity planning that sovereign AI deployment requires.

\subsection{Policy implications}

The FSOR analysis yields three actionable insights for policymakers and infrastructure operators.

First, \textbf{infrastructure investment precedes sovereign AI capacity}. Expanding a region's FSOR requires investment in clean energy interconnection, advanced cooling systems that reduce water intensity, and optical network infrastructure that reduces propagation delay to low-carbon sites. Compute hardware investment alone does not expand the FSOR if the binding constraints are environmental or topological.

Second, \textbf{carbon thresholds must be calibrated to the regional grid baseline}. Setting a carbon policy threshold below the prevailing grid intensity eliminates all local sites from the feasible set, reducing sovereignty to zero regardless of compute capacity.  Effective policy sets thresholds that create credible incentives for grid decarbonization while
maintaining sufficient operational feasibility to sustain AI deployment during the transition.

Third, \textbf{sovereignty is dynamic}. The FSOR changes continuously with grid carbon mix, seasonally with water availability, and over longer timescales with regulatory frameworks. Continuous telemetry and periodic FSOR computation
are therefore necessary to track and manage sovereign AI capacity over time. A region whose FSOR is adequate today may face binding constraints within a planning horizon, depending on grid evolution and climate trends.

\section{Discussion}
\label{sec:discussion}

This section examines the implications of the proposed formulation, emphasizing the structural roles of joint optimization, constraint modeling, and system limitations in sustainability-constrained AI infrastructure. Taken together, these elements show that such systems are not optimization problems with additional costs, but constrained feasibility problems whose solution space is jointly determined by physical infrastructure, environmental limits, and network topology.

\subsection{Why joint optimization matters}

The central finding of this work is that sustainability constraints are intrinsically coupled with both compute placement and network routing, and that this coupling \emph{defines feasibility} rather than merely influencing performance. The consequences of ignoring this coupling are not merely suboptimal solutions: they are solutions whose reported feasibility is illusory because they assume away the network cost that compliance actually incurs.

The coupling arises from a geographic asymmetry: low-carbon and low-water sites are not uniformly distributed. As illustrated in
Figure~\ref{fig:green_but_far}, they are located in specific
regions whose positions within the optical transport network
impose specific propagation delays and capacity constraints on
any workload placed there. This asymmetry is well documented empirically. Grid carbon intensity varies by factors of two or more across cloud regions and exhibits strong spatio-temporal structure~\cite{Sukprasert2024}, making the identity of the lowest-carbon site at any given telemetry cycle both time-varying and location-dependent. As green energy transitions advance, this concentration is likely to intensify rather than dissipate: renewable-heavy grids tend to be located in specific geographic corridors---northern Europe for wind and hydro, southern Europe and North Africa for solar---rather than uniformly distributed~\cite{Worman2024,Koronen2019}.

The implication is that complying with a carbon or water budget is not a free action in the placement space: it restricts workloads to a geographic subset of sites that are, on average, farther from one another and from network aggregation points. Every unit of environmental compliance therefore carries an implicit network cost in the form of additional propagation delay, consumed link capacity, or reduced routing flexibility.

This coupling cannot be resolved by optimizing each dimension independently. Consider two natural decompositions.

A \emph{placement-first} approach selects the site that minimizes the sustainability metric and subsequently solves the routing problem on the resulting fixed placement. This fails because the routing problem inherits a fixed source node that may be poorly positioned relative to the network topology, producing latency values that exceed the hard constraint. The placement appeared feasible in the placement stage because network cost was implicitly assumed to be zero.

A \emph{routing-first} approach minimizes end-to-end latency over all candidate paths and placements without regard to sustainability, then checks whether the selected placement satisfies environmental limits. This produces solutions that are latency-optimal but may violate hard carbon or water caps. Indeed, the FORTE framework, a widely cited baseline for geographic load balancing across data centers, explicitly omits the energy consumed by network transport in its sustainability accounting~\cite{Gao2012,GLDP2016}, a gap identified as a systematic bias in the literature: sequential approaches that ignore network cost underestimate the true environmental and operational burden of spatial workload shifts~\cite{Sukprasert2024,CarbonEndToEnd2024}.

Only joint optimization, simultaneous assignment of workload placement and routing path subject to sustainability and latency constraints, correctly identifies configurations that are simultaneously feasible on both dimensions. In this formulation, the feasible sovereign operating region emerges as the joint intersection of environmental and network constraints, rather than as a post hoc filter applied to independently optimized solutions. This is a classical argument in network optimization: the joint placement and routing problem is generally not decomposable without loss of optimality because placement decisions and routing decisions share link capacity constraints~\cite{Qin2024,JASPER2017}.

This observation has a direct implication for the growing literature on carbon-aware computing. The dominant paradigm in that literature operates at the compute-placement level: it selects sites or time windows based on grid carbon intensity while treating network routing as either free or exogenous~\cite{Sukprasert2024,radovanovic2021carbon,Souza2023CASPER}. As carbon constraints tighten and compliant sites become more geographically concentrated, the network cost of each spatial shift grows. A compute-only carbon accounting framework systematically underestimates this cost. End-to-end carbon footprint analysis of data movement, including the carbon embodied in network transport, confirms that the communication component is non-trivial and grows with the geographic distance of the spatial shift~\cite{CarbonEndToEnd2024}.

This systematic underestimation has a further structural consequence. The practical upper bound on carbon reduction achievable through spatial shifting is already significantly lower than the theoretical ideal due to capacity constraints, latency SLOs, and regional policy restrictions~\cite{Sukprasert2024}. Ignoring network cost makes this bound appear more achievable than it is, producing overconfident estimates of the feasible sovereign operating region.

In this sense, joint optimization is not only a performance improvement over sequential approaches, it is a requirement for correctly identifying whether a feasible solution exists at all. In our formulation, the feasible sovereign operating region is defined by the intersection of the carbon constraints, the water constraints, the latency bound, and the link capacity constraints. This intersection is computed simultaneously over placement and routing variables: it cannot be evaluated by solving the constraints sequentially over separate variable spaces. The infeasibility events of Scenario~C illustrate this precisely: it is the simultaneous binding of environmental and network constraints that exhausts the feasible region, not the violation of any single dimension in isolation.

As the energy transition progresses and low-carbon resources become increasingly geographically concentrated, the interaction between sustainability and network constraints will become more binding. Joint optimization is therefore the minimal formulation required to capture the physical and operational structure of sustainability-constrained AI infrastructure.

\subsection{Hard constraints versus penalty terms}
\label{subsec:hard_vs_penalty_disc}

The choice to model sustainability limits as hard constraints rather than as penalty terms in the objective function is a central design decision of this work and has significant formal and operational consequences. Understanding this distinction requires a brief characterization of both approaches.

In penalty-based formulations, constraint violations are incorporated into the objective function as weighted cost terms, effectively converting a constrained
problem into an unconstrained one~\cite{Nocedal2006,Luenberger2016}. The general idea is to replace hard constraints by penalties and then exploit the well-developed machinery for unconstrained optimization~\cite{Lange2013}.
As the penalty coefficient tends to infinity, the penalized solution converges to a solution of the original constrained problem under standard regularity
conditions~\cite{Luenberger2016}. This approach has computational advantages---it does not restrict the feasible region and can be effective when the constraint surface is non-convex or disconnected---but it carries a fundamental semantic limitation: the optimizer is free to trade constraint violation for objective improvement. In the context of carbon and water budgets, a penalty-based formulation allows the solver to place workloads at sites that breach regulatory thresholds provided this is compensated by sufficiently large performance gains elsewhere in the objective. Solutions returned may therefore satisfy the combined weighted objective while violating absolute regulatory limits under adverse grid or hydrological conditions.

Hard constraints, by contrast, encode the legal and physical reality that certain carbon or water levels are not merely undesirable, they are impermissible. Chinneck~\cite{Chinneck2008} characterizes the fundamental distinction precisely: hard constraints set conditions on the variables that are \emph{required} to be satisfied, whereas soft constraints impose only a preference whose violation is penalized but tolerated. For operators subject to emissions reporting obligations, energy-use disclosure mandates, or regional water-abstraction permits, this distinction is not a modeling preference, it is a structural requirement of regulatory compliance. The energy system optimization literature has increasingly recognized this principle: \citet{Moret2019} note that enforcing constraint feasibility before optimizing cost is the appropriate hierarchical priority in regulatory planning contexts, and robust optimization frameworks for energy systems explicitly distinguish between constraints that must hold under all realizations of uncertainty and those that represent operational targets which may be traded~\cite{Bertsimas2004,Moret2019}.

The cost of encoding sustainability limits as hard constraints is reduced feasibility under extreme conditions. When grid carbon intensity spikes across all candidate sites simultaneously, or when water-stress events raise the effective water intensity beyond per-site policy limits, the intersection of the carbon, water, latency, and capacity constraints may become empty. This is precisely the class of infeasibility events observed in Scenario~C. In the mixed-integer programming literature, such events correspond to what is formally termed an \emph{Irreducible Infeasible Set} (IIS), a minimal subset of constraints that is collectively infeasible, such that the removal of any single constraint restores feasibility~\cite{Chinneck1991,Chinneck2008}. In
our model, the IIS under Scenario~C is characterized by the simultaneous binding of the carbon cap, the water cap, and the latency bound across all candidate sites, leaving no feasible placement for the affected workloads. Detecting and reporting the binding constraints that constitute this IIS is operationally valuable: it identifies the minimum regulatory relaxation or infrastructure augmentation---additional renewable capacity, emergency water allocation, or latency budget extension---that would restore feasibility.

The benefit of the hard-constraint design is that every returned solution is \emph{guaranteed} to be policy-compliant by construction. This guarantee is not replicable in the penalty regime without driving the penalty coefficient to impractically large values, at which point the penalized problem becomes numerically ill-conditioned~\cite{Nocedal2006}. The distinction between feasibility and near-feasibility is material in a regulatory context: a solution that violates a carbon cap by a quantity below the penalty threshold may nonetheless trigger enforcement action, financial penalties, or operating license suspension. Recent data-center scheduling work has reinforced this principle, distinguishing sharply between hard constraints, enforced via constraint projection or shielding mechanisms, and soft constraints addressed by reward shaping or Lagrangian relaxation, with the former reserved for safety and compliance boundaries that cannot be compromised~\cite{Souza2023CASPER}.

A related consideration concerns the treatment of uncertainty. In a deterministic formulation with hard constraints, parameter uncertainty can generate infeasibility when realized values exceed the modeled bounds. Robust optimization offers an intermediate position: constraints are required to hold for all realizations of uncertain parameters within a defined
uncertainty set, trading solution cost for robustness~\cite{BenTal1999, Bertsimas2004}. Applied to carbon intensity, a robust hard constraint would require the carbon budget to be satisfied under the worst-case intensity realization in a defined forecast interval. This is a natural extension of the present model and is identified as a direction for future work.

In summary, the hard-constraint design reflects the legal and operational reality of regulated infrastructure: certain limits are not objectives to be balanced but boundaries to be respected. The infeasibility events that result from this choice are not model failures, they are informative signals that the feasible sovereign operating region has been exhausted under the prevailing conditions, and that the appropriate response is capacity expansion, policy revision, or workload shedding rather than regulatory trade-off. In this sense, infeasibility is not a failure of the optimization process, but a precise indication that the system is operating outside its feasible sovereign operating region.

\subsection{Limitations and future work}
\label{subsec:limitations}

The framework presented in this work makes several simplifying assumptions that bound its scope. Each limitation points to a concrete direction for future work.
The model treats workload placement as an instantaneous decision: once the optimizer assigns a workload to a site, no overhead is attributed to the act of relocation. In practice, relocating a large distributed training job across data centers incurs non-trivial energy and time overhead. Live migration of virtual machines or containers consumes energy at the source host, the destination host, and the intervening backbone network nodes; and the migration time introduces a service disruption window whose duration depends on the memory footprint and dirty-page rate of the workload~\cite{Clark2005}. For training workloads with tight checkpoint synchronization across accelerators, even a brief migration-induced pause may invalidate in-flight gradient aggregation. A complete operational framework should account for migration cost as a function of workload size, page dirty rate, and inter-site link bandwidth, and incorporate a migration feasibility check that prevents relocation when overhead would exceed the sustainability benefit of the destination site. The literature on cost-aware live migration provides the building blocks for such an extension~\cite{Voorsluys2009}.

The model assumes that marginal carbon intensity is observable via real-time telemetry at each site. This assumption may not hold uniformly. Carbon intensity signals are not directly measurable quantities: they are estimated from grid dispatch data, fuel-mix reports, and marginal generator identification, using models that are inherently probabilistic and depend on the transparency and temporal resolution of data published by transmission system operators~\cite{Lindberg2021}. In regions with limited grid transparency---much of sub-Saharan Africa, parts of South and Southeast Asia, and several emerging cloud markets---providers must fall back on synthetic proxy models trained on historical load patterns, introducing estimation errors that can be substantial~\cite{wiesner2025}. Even in well-instrumented grids, the choice between average and marginal emissions factors has a large effect on the signal used for placement decisions~\cite{ElectricityMaps2023}: average factors incentivize shifting load to times when the overall mix is clean, while marginal factors target the actual incremental emitter, and the two can point in opposite directions during periods of high renewable curtailment~\cite{wiesner2025}. A robust extension of the present framework should model carbon intensity as an uncertain parameter drawn from a forecast distribution, replacing the deterministic hard constraint with its robust counterpart as outlined in~\S\ref{subsec:hard_vs_penalty_disc}, and evaluate the sensitivity of placement decisions to signal quality and temporal resolution.

The exact MILP formulation is tractable for the scenario sizes considered in this work---a modest number of sites, workload classes, and telemetry cycles. Scaling to continental-scale deployments with hundreds of candidate sites, fine-grained routing topologies, and rolling time horizons requires more scalable solution methods. Benders decomposition offers a natural structural decomposition: the placement master problem, which assigns workloads to sites subject to sustainability constraints, and the routing subproblem, which solves the min-latency path given a fixed placement, are coupled only through site assignment variables and link capacity constraints~\cite{Benders1962,Rahmaniani2017}. This block structure permits the application of standard Benders cuts to iteratively tighten the placement problem without solving the full joint formulation~\cite{Bonami2020}. For very large instances, Lagrangian relaxation of the coupling constraints provides a scalable lower bound that can guide heuristic placement policies~\cite{Fisher1981}. An alternative route to scalability is learning-based approximation: recent work has demonstrated that graph neural networks can learn to predict near-optimal placement decisions from problem features, reducing online solve time by orders of magnitude at the cost of a modest optimality gap~\cite{Cappart2023}. The appropriate scalability method depends on whether the operator requires provable feasibility guarantees, which Benders provides, or is willing to accept a feasibility certificate by inspection in exchange for faster response.

The framework addresses operational placement decisions under fixed infrastructure: the set of candidate sites, their power headroom, and the optical link capacities are treated as exogenous parameters. It does not model the longer-term capital investment decisions that determine the FSOR itself, new data center construction, renewable energy procurement contracts, optical capacity expansion, and grid interconnection agreements. These decisions operate on timescales of years to decades and interact with operational decisions in both directions: operational infeasibility events (such as those of Scenario~C) signal that the current infrastructure endowment cannot support the workload portfolio under the prevailing sustainability limits, providing an investment signal; conversely, capital investment decisions reshape the feasible sovereign operating region that operational planning must navigate. Integrating operational and investment planning into a two-stage or multi-period stochastic program represents a natural and important direction for future work~\cite{Moret2019,Reinert2023}. The growing literature on joint data center siting and energy procurement planning under carbon objectives provides relevant methodologies~\cite{Koronen2019,Sukprasert2024}.

More broadly, the framework assumes that infrastructure resources are available and controllable by the operator subject to the physical and regulatory constraints modeled. This assumption may not hold in environments with strong external dependencies: a workload placed on infrastructure operated by a third-party cloud provider is subject to that provider's scheduling policies, capacity allocation priorities, and jurisdictional obligations, which may override the operator's sustainability and sovereignty requirements~\cite{Daskal2018}. Incorporating such dependency constraints into the feasibility model requires extending the notion of a feasible sovereign operating region to account not only for physical and regulatory limits but also for contractual and legal boundaries on what the operator can unilaterally control. Formalizing this extension---mapping the space of actions available under different infrastructure ownership and contractual structures---represents an important direction for future work at the intersection of infrastructure optimization and digital sovereignty.

\section{Conclusion}
\label{sec:conclusion}

This paper introduced a sustainability-constrained workload
orchestration framework for AI infrastructure. By treating carbon
intensity, water availability, and power capacity as hard feasibility constraints rather than optimization objectives, the framework captures the regulatory and physical reality that certain environmental conditions are impermissible---not merely undesirable. By solving compute placement and optical network routing jointly, it resolves the green-but-far tension---the structural conflict between sustainability eligibility and latency admissibility---that decoupled approaches cannot represent as a constraint interaction.

The Feasible Sovereign Operating Region provides a concrete,
telemetry-grounded metric for sovereign AI capacity. Rather than
measuring sovereignty through data localization or system ownership, the FSOR quantifies the set of workloads that a given infrastructure can actually sustain under its physical and regulatory endowment. Scenario-based analysis shows that joint optimization consistently outperforms both unconstrained and compute-only baselines on environmental impact, while preserving latency guarantees for latency-sensitive workloads. For the scenario sizes considered, all problem instances were solved to certified optimality within a five-minute telemetry cycle using an open-source MILP solver, confirming operational tractability at these scales. Infeasibility events under water-stress conditions demonstrate that sustainability
limits manifest as hard operational boundaries, providing operators with precise signals about the conditions under which sovereign AI operation is and is not feasible.

These results reinforce the thesis that sovereignty in AI
infrastructure is not a software property. It is an engineering
outcome determined by the physical, environmental, and topological characteristics of the underlying system. Expanding sovereign AI capability requires coordinated investment in clean energy, efficient cooling, and optical connectivity---alongside, but not replaced by, advances in compute technology. A region that neglects its infrastructure endowment cannot compensate through software optimization. The FSOR makes that boundary visible.

\bibliographystyle{elsarticle-num}

\bibliography{biblio}

\section*{Author biography}\noindent

\bioItem[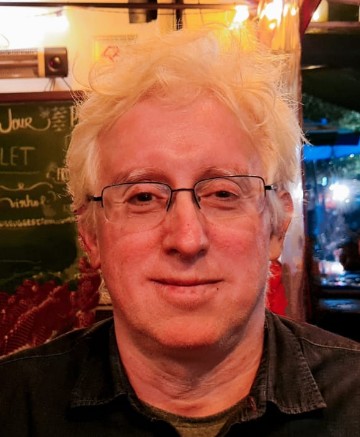]{Sergio Cruzes}{ received his MSc.\ degree
in electrical engineering from the University of S{\~{a}}o Paulo
(EESC) with focus on optical communications, where he developed a
prototype for video transmission over fiber optics using frequency
modulation. He received his MBA degree in project management and
technological innovation from FIPE Brazil in 2010. He currently
works as an optical network engineer at Ciena Brazil, where his
work spans optical transport, network automation, and the
intersection of AI infrastructure with sustainable operations.
He is the author of the IEEE Access paper ``Telemetry and Agentic
AI: Foundations for Optical Network Automation,'' the IEEE paper
``Failure Management Overview in Optical Networks,'' the
\textit{Optical Switching and Networking} paper ``Revolutionizing
Optical Networks: The Integration and Impact of Large Language
Models,'' and the preprint ``AI Infrastructure Sovereignty,''. His research interests include optical network automation, agentic AI architectures for network control,
large language model applications in telecommunications, modern
data center infrastructure and its physical constraints, AI
infrastructure sovereignty and the role of environmental and
topological endowments in determining regional AI capacity,
high-spectral-efficiency coherent transmission systems,
disaggregated optical networks, quality of transmission
estimation, and failure management in optical networks.}
\printBio
\end{document}